\documentclass[11pt]{article}
\usepackage{amssymb}
\usepackage{graphicx}
\usepackage{amsmath}
\usepackage{makeidx}
\usepackage{indentfirst}
\usepackage[T1]{fontenc}
\usepackage[utf8]{inputenc}
\usepackage{authblk}

\newcounter{resultnum}[section]
\newcounter{conclusionnum}[section]
\newcounter{conditionnum}[section]
\newcounter{conjecturenum}[section]
\newcounter{examplenum}[section]
\newcounter{exercisenum}[section]
\newcounter{lemmanum}[section]
\newcounter{notationnum}[section]
\newcounter{theoremnum}[section]
\newcounter{definitionnum}[section]
\newcounter{corollarynum}[section]
\newcounter{remarknum}[section]
\newcounter{propositionnum}[section]
\newcounter{acknowledgementnum}[section]
\newcounter{algorithmnum}[section]
\newcounter{axiomnum}[section]
\newcounter{casenum}[section]
\newcounter{claimnum}[section]
\newcounter{summarynum}[section]
\newcounter{problemnum}[section]
\begin{document}

\title{Cosmological Attractors and Anisotropies in Two Measure Theories, Effective EYMH systems, and Off--Diagonal Inflation Models}
\date{May 4, 2017}
\author{ 
{ \textbf{Subhash Rajpoot}}\\
{\small \textit{California State University at Long Beach, }}%
{\small \textit{Long Beach, California, USA }}\\
{\small \textit{email: Subhash.Rajpoot@csulb.edu ${}$ }}\\
\vspace{.2 in} \textbf{Sergiu I. Vacaru}\footnote{{\it Address for correspondence:\ } Flat 4, Brefney house, Fleet street, Ashton-under-Lyne, Lancashire, OL6 7PG, the UK }\\
{\small \textit{Quantum Gravity Research; 101 S. Topanga Canyon Blvd \#
1159. Topanga, CA 90290, USA}} \\
{\small and } \ {\small \textit{ Project IDEI, University "Al. I. Cuza" Ia\c si, Romania}}\\
{\small \textit{email: sergiu.vacaru@gmail.com }}}

\maketitle

\begin{abstract}
Applying the anholonomic frame deformation method, we construct various classes of cosmological solutions for effective Einstein -- Yang-Mills -- Higgs,  and two measure theories.  The types of models considered are Freedman-Lema\^{i}tre-Robertson-Walker, Bianchi, Kasner and models with  attractor configurations.  The various regimes pertaining to  plateau--type inflation,  quadratic inflation,  Starobinsky type and Higgs type inflation are presented.
\vskip0.2cm

\vskip0.2cm

\vskip0.2cm

\end{abstract}

\renewcommand\Authands{and }

\tableofcontents

\section{Introduction }
\label{s1}
Over time, the Cosmological Constant Problem (CCP) has evolved from the "Old Cosmological Constant Problem" \cite{W1},  where  the concern was on  why the observed vacuum energy density of the universe is exactly zero, to the present form pertaining to the evidence establishing the accelerating expansion of the universe  \cite{AU}.  One is therefore  faced with  the   "New Cosmological Constant Problem" \cite{W2,sahni}.  In other words, the  problem has shifted  from not why the CCP is  exactly zero, but  to why the  vacuum energy density is so small. Various attempts to address the issue  range from the conventional to the esoteric. Conventional field theoretic models are based on a single scalar field (quintessence) while the  esoteric models involve tachyons, phantoms and K-essence. The latter may also admit  multi scalar field configurations.
Such models have also been supplemented further to take into consideration the recent observational
data from Planck \cite{planck1,planck2} and BICEP2 \cite{bicep2}.  In all these models the inflationary paradigm \cite{guth,linde1} is the underlying theme, see also an opposite point of view in \cite{ijj}. However, present data is insufficient to determine precisely what the initial conditions  were  that drove inflation. In addressing the present situation there are essentially two main approaches entertained.  In one approach it is assumed that  there is a basic mechanism driving to  zero the vacuum energy but some "residual" interactions survive that slightly
shift the vacuum energy density towards the presently observed  small non--zero value. In the alternative approach it is assumed that the true vacuum energy will exactly be  zero when  the final state of the   theory  is reached and the present state pertaining to the small non zero vacuum energy density is the result of our universe having not reached that final state  yet. \\
\indent In this work, we will adapt the view point that the above two scenarios represent equally viable solutions to the CCP and both can  be entertained  naturally if one considers off--diagonal inhomogeneous cosmological solutions.
 Alternative constructs are also possible and are discussed in \cite%
{linde1,linde2,linde3}  in a different class of theories. As will be demonstrated, for certain well defined conditions, the  models considered in this work can be treated as effective two measure theories (TMTs) studied in Refs. \cite%
{guend0,guend1,mli,guend2,guend5,guend7}. In these theories, the  modified gravitational and matter field equations of TMTs generate effective Einstein - Yang-Mills - Higgs (EYMH) systems which can be solved in analytic form using geometric methods. The underlying principle of the geometric method is based on  the anholonomic frame deformation method (AFDM) \cite{vhep,vsingl1,vcosms,vnb15,vhd2013}. The main idea of  the AFDM is to re--write equivalently Einstein equations, and various modifications of it, on a (pseudo) Riemannian  manifold $\mathbf{V}$ in terms of an "auxiliary" linear connection $\mathbf{D}$.  This connection, together with the
Levi--Civita (LC) connection $\nabla $,  is defined in a metric compatible form  by a split metric structure $\mathbf{g}=\{\mathbf{g}_{\alpha \beta }=[g_{ij},g_{ab}]\}.$ In order to establish our notation,
 we take $\dim \mathbf{V}=4,$ with  the conventional splitting of coordinates as  $3+1$, and the
equivalent splitting as $2+2$ respectively. The signature of the metric on $\mathbf{V%
}$ is taken to be  $(+,+,+,-).$ Indices $i,j,k,...$ take values $1,2$ while indices $%
a,b,...$ take values $3,4$ and the local coordinates are denoted by  $u^{\alpha
}=(x^{i},y^{a}),$ or collectively as  $u=(x,y).$\footnote{The 2+2 splitting is convenient for constructing exact cosmological solutions with generic off-diagonal metrics which can not be diagonalized by coordinate transforms in a finite spacetime region. Nevertheless, realistically,  we shall have to consider 3+1 splitting, for instance, in section \ref{ssoddflrw} in order to study off-diagonal deformations of FLRW configurations in TMTs, with effective fluid energy-momentum stress tensor. } Quantities under consideration and with a left label (for
instance, $\ ^{\mathbf{g}}\mathbf{D}$ )  emphasize that the
geometric object ($\mathbf{D}$)  is uniquely determined by $\mathbf{g}$. Unless otherwise stated, Einstein's summation convention is assumed throughout with the caveat that
upper and lower labels are omitted if this does not result in ambiguities.  We emphasize  that $%
\mathbf{D}$ contains nontrivial anholonomically induced torsion $\mathbf{T}$
relating  to the underlying nonholonomic frame structure.
Such a torsion field is completely defined by the metric and the
nonholonomic (equivalently, anholonomic and/or non--integrable) distortion relations,
\begin{equation}
\ \mathbf{D}=\ \nabla +\mathbf{Z}[\mathbf{T}],  \label{distr}
\end{equation}%
when both the linear connections and the distortion tensor $\mathbf{Z}[\mathbf{T}%
]$ are uniquely determined by certain well--defined geometric and/or physical
principles. Physical models are constructed following the
principle that all geometric constructions are adapted to a nonholonomic
splitting with an associated nonlinear connection (N--connection) structure $%
\mathbf{N}=\{N_{i}^{a}(u)\}$ that splits into the  Whitney sum consisting of the conventional horizontal (h) and vertical
(v) components,
\begin{equation}
\mathbf{N}:\ T\mathbf{V}=\ ^{h}\mathbf{V\oplus }\ ^{v}\mathbf{V}\equiv h%
\mathbf{\mathbf{V\oplus }}v\mathbf{\mathbf{V},}  \label{ncon}
\end{equation}%
where $T\mathbf{V}$ is the tangent bundle\footnote{Boldface symbols will be used in
order to emphasize that certain spaces and/or geometric objects are
adapted to a N--connection. Here we note that, for instance,  $\ ^{h}\mathbf{V}$ is equivalent to $h\mathbf{V}$ (in order to avoid ambiguities, we present both types of notations used in our former  works and references therein). Such a conventional decomposition (equivalently, fibred structure) can  always be constructed on any 4-d metric-affine manifold. In general relativity, it is known as the diadic decomposition of tetrads. The most important outcome  of our works \cite{vhep,vsingl1,vcosms,vnb15,vhd2013} is that we proved that (modified) Einstein equations can be decoupled and solved in very general forms both  for a N--adapted 2+2 splitting and a d-connection $\widehat{\mathbf{D}}$ (this auxiliary connection was not considered in former works with diadic structures). }. For such a
splitting, all geometric constructions can be carried out equivalently with $%
\nabla $  using the so--called canonical distinguished connection
(d--connection), $\widehat{\mathbf{D}}$. Here  $\widehat{\mathbf{D}}$  is distinct from $\mathbf{D}$. This linear connection is
N--adapted, i.e. preserves under parallelism the N--connection splitting, and is uniquely determined (together with $\nabla $) by the constraints
\begin{equation}
\mathbf{g}\rightarrow \left\{
\begin{array}{ccccc}
\nabla : &  & \nabla \mathbf{g}=0;\ ^{\nabla }\mathbf{T}=0, &  &
\mbox{ the
Levi--Civita connection;} \\
\widehat{\mathbf{D}}: &  & \widehat{\mathbf{D}}\ \mathbf{g}=0;\ h\widehat{%
\mathbf{T}}=0,\ v\widehat{\mathbf{T}}=0, &  &
\mbox{ the canonical
d--connection.}%
\end{array}%
\right.   \label{cdcon}
\end{equation}%
It is to be noted that in  general, a d--connection $\mathbf{D}$ can  equivalently split into the N--adapted
horizontal (h) and vertical (v) components, respectively, as $h{\mathbf{D}}$ and $v{\mathbf{D}}$, (or equivalently, as $=(^h{{\mathbf{D}}},^v{{\mathbf{D}}})$). But such a splitting may  not be compatible, (i.e., ${\mathbf{D}}\ \mathbf{g}\neq 0,$ ) as it can carry arbitrary amount of torsion $%
\mathbf{T}$, and hence is not  subject to the aforementioned constraints depicted in (\ref%
{cdcon}).

The advantage of the canonical d--connection $\widehat{\mathbf{D}}$ is that in this framework hatted Einstein equations result,
\begin{equation}
\widehat{\mathbf{G}}_{\alpha\beta}:=\widehat{R}_{\alpha \beta }-\frac{1}{2}\mathbf{g}_{\alpha\beta }\ \widehat{R}=\mathbf{\Upsilon}_{\alpha\beta}(u).  \label{nceq}
\end{equation}%
 Here the hatted Einstein tensor $\widehat{\mathbf{G}}$ and the effective source term $%
\mathbf{\Upsilon}$ are defined in standard form following geometric methods and
N--adapted variational calculus but for quantities  $(\mathbf{g},\widehat{\mathbf{D}})$ instead of the usual $(\mathbf{g},\nabla )$. The hatted Einstein equations decouple with respect to a class of N--adapted frames for various classes of
metrics with one--Killing symmetry \cite{vcosms,vnb15}.
   This allows us to integrate (\ref%
{nceq}) in a very general form by generic off--diagonal metrics, metrics
that otherwise can not be diagonalized in a finite spacetime region by coordinate
transformations  that are determined via a set of generating and integration functions depending on
all spacetime coordinates and various types of commutative or noncommutative parameters\footnote{In general, symmetric metrics of the type $\mathbf{g}_{\alpha
\beta }(x^{1},x^{2},y^{3},y^{4}=t),$ with $t$ being a timelike coordinate,
contain a maximum of six independent variables since  four coefficients from the ten
components of the metric tensor of a 4--d spacetime can be transformed away via
coordinate transforms as a result of  the Bianchi identities. }. Solutions thus determined
describe various geometric and physical models in modified gravity theories with
nontrivial nonholonomically induced torsion, $\widehat{\mathbf{T}}\neq 0,$
and generalized connections. As special cases, we  extract LC--configurations and construct
new classes of cosmological solutions in Einstein's gravity if
we constrain the set of possible generating and integration functions to
satisfy the following conditions,
\begin{eqnarray}
\widehat{\mathbf{T}} &=&0,  \label{lc} \\
\mathbf{g}_{\alpha \beta } &=&\mathbf{g}_{\alpha \beta }(t),  \label{cosmc}
\end{eqnarray}%
where metric $\mathbf{g}_{\alpha \beta }(t)$  in the unprimed  bases can be related to metric in the primed bases via frame transformations, i.e., $\mathbf{g}_{\alpha \beta }(t)=e_{\ \alpha
}^{\alpha ^{\prime }}e_{\ \beta }^{\beta ^{\prime }}\mathbf{g}_{\alpha
^{\prime }\beta ^{\prime }}(t),$ where  \ $e_{\ \alpha}^{\alpha ^{\prime }}$ represents  the tetrad frame field. For instance, $\mathbf{g}_{\alpha ^{\prime }\beta
^{\prime }}$ can be a Bianchi type  metric, or a diagonalized
homogeneous  Friedmann--Lema\^{\i}tre--Robertson--Walker (FLRW)
type metric. In general, $\mathbf{g}_{\alpha ^{\prime }\beta ^{\prime }}$ may
not be a solution of any gravitational field equations but we shall always impose
 the constraint that it's nonholonomic deformation $\mathbf{g}_{\alpha
\beta }$  always is a solution of the hatted Einstein equation(\ref{nceq}).

In general, gravitational field equations (\ref{nceq}) constitute a
sophisticated system of nonlinear partial differential equations (PDEs) as opposed to the occurrence of ordinary differential equations (ODEs) in conventional general relativity. The AFDM, on the other hand, allows us to find new classes of solutions by decoupling the PDEs.
  We emphasize that in the AFDM approach advocated here,
constraints of type (\ref{lc}) and/or (\ref{cosmc}) are to be imposed after
the inhomogeneous $\mathbf{g}_{\alpha \beta }(x^{i},y^{3},t)$ are constructed
in general form. If the aforementioned constraints are imposed from the very beginning in
order to transform PDEs into ODEs,  a large class of generic
off--diagonal and diagonal solutions will be compromised.  The
specific goal of this work is to apply the AFDM method and explicitly construct  solutions  in effective TMTs addressing attractors, acceleration, dark energy and dark matter effects in the new cosmological models.

This work is organized as follows. In section \ref{s2}, we provide a brief
introduction to the geometry of nonholonomic deformations in Einstein
gravity and modifications that lead to effective TMTs. In such theories we  shown how the
gravitational and matter field equations  can be decoupled
and solved in very general off--diagonal forms for the canonical
d--connection with constraints for LC--configurations. Section \ref{s3} is
devoted to off--diagonal and diagonal cosmological solutions with small
vacuum density. Also constructed and analysed are the off--diagonal
inhomogeneous cosmological solutions with nonholonomically induced torsion.
  In section  \ref{s4}, we study the equivalence of
effective TMTs with sources for nonlinear potentials and EYMH self--dual
fields resulting in  attractor type behaviour. In  section \ref{s5}, we  analyze in explicit form how
exact cosmological solutions with locally anisotropic attractor properties
can be generated by deforming FLRW type diagonal metrics and off--diagonal
Bianchi type cosmological models. Conclusions are presented in section \ref{s6}.

\section{Nonholonomic Deformations}
\label{s2}

For clarity, we elaborate upon  our notation first. On a (pseudo) Riemannian manifold we prescribe an N--connection with horizontal($h$) and vertical($v$) decompositions ($h$ and $v$ splitting) (\ref{ncon})as $(\mathbf{V},\ \mathbf{N})$. To this we  associate
structures of N--adapted\ local bases, $\mathbf{e}_{\nu }=(\mathbf{e}%
_{i},e_{a}),$ and cobases, $\mathbf{e}^{\mu }=(e^{i},\mathbf{e}^{a}),$ which
are the following  N--elongated partial derivatives and differentials,
\begin{eqnarray}
&&\mathbf{e}_{i} :=\partial /\partial x^{i}-\ N_{i}^{a}(u)\partial /\partial
y^{a},\ e_{a}:=\partial _{a}=\partial /\partial y^{a},  \label{nader} \\
&&\mbox{\qquad and  }e^{i} =dx^{i},\ \mathbf{e}^{a}=dy^{a}+\
N_{i}^{a}(u)dx^{i}.  \label{nadif}
\end{eqnarray}%
The frame basis $\mathbf{e}_{\nu }=(\mathbf{e}_{i},e_{a}),$
satisfy the nonholonomy relations
\begin{equation}
\lbrack \mathbf{e}_{\alpha },\mathbf{e}_{\beta }]=\mathbf{e}_{\alpha }%
\mathbf{e}_{\beta }-\mathbf{e}_{\beta }\mathbf{e}_{\alpha }=W_{\alpha \beta
}^{\gamma }\mathbf{e}_{\gamma },  \label{nonholr}
\end{equation}%
with nontrivial nonholonomy coefficients
\begin{equation}
W_{ia}^{b}=\partial _{a}N_{i}^{b}, \,\,\,\,\, W_{ji}^{a}=\Omega _{ij}^{a}=\mathbf{e}%
_{j}\left( N_{i}^{a}\right) -\mathbf{e}_{i}(N_{j}^{a}).  \label{anhcoef}
\end{equation}%
Such a basis is holonomic if and only if $W_{\alpha \beta }^{\gamma }=0.$
This is trivially satisfied in a coordinate
basis if  $\mathbf{e}_{\alpha }=\partial _{\alpha }$ . As holonomic dual basis, we take  $\mathbf{e}^{\mu }=du^{\mu }$.

The geometric objects on $\mathbf{V}$ are defined with respect to the N--adapted frames (\ref{nader}),
(\ref{nadif}). These are referred to as  distinguished objects or d--objects in short.
A vector $Y(u)\in T\mathbf{V}$ is parameterized as a
d--vector. Explicitly,  $\mathbf{Y}=$ $\mathbf{Y}^{\alpha }\mathbf{e}_{\alpha }=\mathbf{Y}%
^{i}\mathbf{e}_{i}+\mathbf{Y}^{a}e_{a},$ or $\mathbf{Y}=(hY,vY),$ with $hY=\{%
\mathbf{Y}^{i}\}$ and $vY=\{\mathbf{Y}^{a}\}.$ Likewise, in this frame work,  the coefficients of d--tensors, N--adapted differential forms, d--connections, and d--spinors are easily accommodated.

Any metric tensor $\mathbf{g}$ on $\mathbf{V}$, defined as a second rank symmetric tensor, takes the following structure with respect to the dual local coordinate basis,
\begin{equation*}
\mathbf{g}=\underline{g}_{\alpha \beta }du^{\alpha }\otimes du^{\beta },
\end{equation*}%
where
\begin{equation}
\underline{g}_{\alpha \beta }=\left[
\begin{array}{cc}
g_{ij}+N_{i}^{a}N_{j}^{b}g_{ab} & N_{j}^{e}g_{ae} \\
N_{i}^{e}g_{be} & g_{ab}%
\end{array}%
\right] .  \label{ansatz}
\end{equation}%
Equivalently,  $\mathbf{g}$  serves as the  d--metric and in tensor product  notation, is taken to be
\begin{equation}
\mathbf{g}=g_{\alpha }(u)\mathbf{e}^{\alpha }\otimes \mathbf{e}^{\beta
}=g_{i}(x)dx^{i}\otimes dx^{i}+g_{a}(x,y)\mathbf{e}^{a}\otimes \mathbf{e}%
^{a}.  \label{dm1}
\end{equation}

Linear connections on $\mathbf{V}$ are introduced in N--adapted and  N--non adapted forms in the standard way. By definition, a d--connection $\mathbf{D}=(hD,vD)$ preserves under parallelism the
N--connection splitting (\ref{ncon}). Any d--connection $\mathbf{D}$ acts as
covariant derivative operator, $\mathbf{D}_{\mathbf{X}}\mathbf{Y}$,
for a d--vector $\mathbf{Y}$ in the direction of a d--vector $\mathbf{X}.$
With respect to N--adapted frames (\ref{nader}) and (\ref{nadif}), we can
compute  the relevant quantities  of interest in N--adapted coefficient form when $\mathbf{D}=\{%
\mathbf{\Gamma }_{\ \alpha \beta }^{\gamma
}=(L_{jk}^{i},L_{bk}^{a},C_{jc}^{i},C_{bc}^{a})\}$. The coefficients $%
\mathbf{\Gamma }_{\ \alpha \beta }^{\gamma }$ are computed for the
horizontal and vertical components of $\mathbf{D}_{\mathbf{e}_{\alpha }}\mathbf{e}_{\beta }:=$
$\mathbf{D}_{\alpha }\mathbf{e}_{\beta }$ by substituting  $\mathbf{X}$ for $\mathbf{e}%
_{\alpha }$ and $\mathbf{Y}$ for $\mathbf{e}_{\beta }.$

We  compute the d--torsion $\mathcal{T},$ the d--torsion nonmetricity $\mathcal{Q},$ and the d--curvature $\mathcal{R}$   for any d--connection $\mathbf{D}$ from the following  standard formulae,
\begin{eqnarray}
&&\mathcal{T}(\mathbf{X,Y}):= \mathbf{D}_{\mathbf{X}}\mathbf{Y}-\mathbf{D}_{%
\mathbf{Y}}\mathbf{X}-[\mathbf{X,Y}],\mathcal{Q}(\mathbf{X}):=\mathbf{D}_{%
\mathbf{X}}\mathbf{g,}  \label{dt} \\
&&\mathcal{R}(\mathbf{X,Y}):= \mathbf{D}_{\mathbf{X}}\mathbf{D}_{\mathbf{Y}}-%
\mathbf{D}_{\mathbf{Y}}\mathbf{D}_{\mathbf{X}}-\mathbf{D}_{\mathbf{[X,Y]}}.
\label{dr}
\end{eqnarray}%
The N--adapted coefficients are correspondingly labeled as
\begin{eqnarray*}
\mathcal{T} &=&\{\mathbf{T}_{\ \alpha \beta }^{\gamma }=\left( T_{\
jk}^{i},T_{\ ja}^{i},T_{\ ji}^{a},T_{\ bi}^{a},T_{\ bc}^{a}\right) \},%
\mathcal{Q}=\mathbf{\{Q}_{\ \alpha \beta }^{\gamma }\}, \\
\mathcal{R} &\mathbf{=}&\mathbf{\{R}_{\ \beta \gamma \delta }^{\alpha }%
\mathbf{=}\left( R_{\ hjk}^{i}\mathbf{,}R_{\ bjk}^{a}\mathbf{,}R_{\ hja}^{i}%
\mathbf{,}R_{\ bja}^{c},R_{\ hba}^{i},R_{\ bea}^{c}\right) \}.
\end{eqnarray*}
The Levi--Civita connection $\nabla $ ( LC) and the canonical
d--connection $\widehat{\mathbf{D}}$ defined by formulas (\ref{cdcon}) are also
expressed   in terms of  the local  N--adapted form.
 The coefficients of $\widehat{\mathbf{D}}=\{\widehat{\mathbf{\Gamma }}_{\ \alpha
\beta }^{\gamma }=(\widehat{L}_{jk}^{i},\widehat{L}_{bk}^{a},\widehat{C}%
_{jc}^{i},\widehat{C}_{bc}^{a})\}$ depend on ($g_{\alpha \beta }$,  $%
N_{i}^{a}$ ) and are computed using the following formulae
\begin{eqnarray}
\widehat{L}_{jk}^{i} &=&\frac{1}{2}g^{ir}\left( \mathbf{e}_{k}g_{jr}+\mathbf{%
e}_{j}g_{kr}-\mathbf{e}_{r}g_{jk}\right) ,\widehat{C}_{bc}^{a}=\frac{1}{2}%
g^{ad}\left( e_{c}g_{bd}+e_{b}g_{cd}-e_{d}g_{bc}\right)  \label{candcon} \\
\widehat{C}_{jc}^{i} &=&\frac{1}{2}g^{ik}e_{c}g_{jk},\ \widehat{L}%
_{bk}^{a}=e_{b}(N_{k}^{a})+\frac{1}{2}g^{ac}\left( \mathbf{e}%
_{k}g_{bc}-g_{dc}\ e_{b}N_{k}^{d}-g_{db}\ e_{c}N_{k}^{d}\right) .  \notag
\end{eqnarray}%
By using  the coefficients of $\nabla =\{\Gamma _{\ \alpha \beta }^{\gamma }\}$,
written with respect to (\ref{nader}) and (\ref{nadif}), we compute the
coefficients of the distortion d--tensor $\widehat{\mathbf{Z}}_{\ \alpha
\beta }^{\gamma }=\widehat{\mathbf{\Gamma }}_{\ \alpha \beta }^{\gamma
}-\Gamma _{\ \alpha \beta }^{\gamma },$ which is the N--adapted coefficient
formula for (\ref{distr}). We elaborate upon geometric and physical models in
equivalent form by working with two metric compatible connections $\widehat{%
\mathbf{D}}$ and $\mathbf{\nabla }$ because all N--adapted coefficients for $%
\widehat{\mathbf{Z}}_{\ \alpha \beta }^{\gamma }=\widehat{\mathbf{\Gamma }}%
_{\ \alpha \beta }^{\gamma }$ and $\Gamma _{\ \alpha \beta }^{\gamma }$ are
completely defined by the same metric structure $\mathbf{g.}$ The nontrivial
d--torsions coefficients\ $\widehat{\mathbf{T}}_{\ \alpha \beta }^{\gamma }$
are computed by setting $\mathbf{D}=\widehat{\mathbf{D}}$ in (\ref{dt})
and determined by the nonholonomy relations,
\begin{equation}
\widehat{T}_{\ jk}^{i}=\widehat{L}_{jk}^{i}-\widehat{L}_{kj}^{i}, \,\,\, \widehat{T}%
_{\ ja}^{i}=\widehat{C}_{jb}^{i}, \,\,\, \widehat{T}_{\ ji}^{a}=-\Omega _{\ ji}^{a},%
\,\,\, \widehat{T}_{aj}^{c}=\widehat{L}_{aj}^{c}-e_{a}(N_{j}^{c}), \,\,\, \widehat{T}_{\
bc}^{a}=\ \widehat{C}_{bc}^{a}-\ \widehat{C}_{cb}^{a}.  \label{dtors}
\end{equation}

Any (pseudo) Riemannian geometry is formulated on a nonholonomic
manifold $\mathbf{V}$ using two equivalent geometric quantities, $(\mathbf{%
g,\nabla })$ or $(\mathbf{g,N,}\widehat{\mathbf{D}}).$ In the
"standard" method we take $\mathbf{D\rightarrow \nabla }$ when $\ ^{\mathbf{%
\nabla }}T_{\ \alpha \beta }^{\gamma }=0,\ ^{\mathbf{\nabla }}Q_{\ \alpha
\beta }^{\gamma }=0$, and $\ ^{\mathbf{\nabla }}R_{\ \beta \gamma \delta
}^{\alpha }$ is computed following formulae (\ref{dr}). For the "geometric
variables" $(\mathbf{g,N,}\widehat{\mathbf{D}})$, using similar formulae, we
compute  $\mathbf{D}=\widehat{\mathbf{D}}$ in standard form
respectively the Riemann  d--tensor $\widehat{\mathcal{R}}$
 and the Ricci  d--tensor $\widehat{\mathcal{R}}ic \{=\widehat{\mathbf{R}}_{\ \beta \gamma }\} $. The nonsymmetric d--tensor $\widehat{\mathbf{R}}_{\alpha \beta }$ of $\widehat{\mathbf{D}}$
is characterized by the following four $h$ and $v$ N--adapted coefficients%
\begin{equation}
\widehat{\mathbf{R}}_{\alpha \beta }=\{\widehat{R}_{ij}:=\widehat{R}_{\
ijk}^{k},\ \widehat{R}_{ia}:=-\widehat{R}_{\ ika}^{k},\ \widehat{R}_{ai}:=%
\widehat{R}_{\ aib}^{b},\ \widehat{R}_{ab}:=\widehat{R}_{\ abc}^{c}\},
\label{driccic}
\end{equation}%
and the  "alternative" scalar curvature
\begin{equation}
\ \widehat{R}:=\mathbf{g}^{\alpha \beta }\widehat{\mathbf{R}}_{\alpha \beta
}=g^{ij}\widehat{R}_{ij}+g^{ab}\widehat{R}_{ab}.  \label{sdcurv}
\end{equation}%
The Einstein d--tensor of $\widehat{\mathbf{D}}$ in hatted form is
\begin{equation}
\widehat{\mathbf{G}}_{\alpha \beta }:=\widehat{\mathbf{R}}_{\alpha \beta }-%
\frac{1}{2}\mathbf{g}_{\alpha \beta }\ \widehat{R}  \label{enstdt}
\end{equation}%
and is a nonholonomic distortion of the standard form, $G_{\alpha \beta }:=R_{\alpha \beta }-\frac{1}{2}\mathbf{g}_{\alpha
\beta }\ R$ , that is computed from   $\mathbf{\nabla }.$
 We solve the equations resulting from the constraints (\ref{lc}) and get solutions to a  system of first order PDE equations
\begin{equation}
\widehat{L}_{aj}^{c}=e_{a}(N_{j}^{c}), \,\,\, \widehat{C}_{jb}^{i}=0,\Omega _{\
ji}^{a}=0.  \label{lccond}
\end{equation}%


Nonholonomic deformations of fundamental geometric objects on a
pseudo--Riemannian manifold $\mathbf{V}$ with N--connection 2+2 splitting
are determined by the transforming of the fundamental geometric data $(\mathbf{%
\mathring{g},\mathring{N},\ }^{\circ }\widehat{\mathbf{D}})\rightarrow (%
\widehat{\mathbf{g}}\mathbf{,N,}\widehat{\mathbf{D}})$, where the "prime"
data $(\mathbf{\mathring{g},\mathring{N},\ }^{\circ }\widehat{\mathbf{D}})$
may or not be a solution of certain gravitational field equations in a (modified) theory of gravity but the "target" data $(\widehat{\mathbf{g}}\mathbf{,N,}\widehat{\mathbf{D}})$ affirmatively define exact solutions of (\ref%
{nceq}) with metrics parameterized in the form (\ref{ansatz}) and (\ref{dm}).

The prime metric is parameterized as
\begin{eqnarray*}
\mathbf{\mathring{g}} &=&\mathring{g}_{\alpha }(u)\mathbf{\mathring{e}}%
^{\alpha }\otimes \mathbf{\mathring{e}}^{\beta }=\mathring{g}%
_{i}(x)dx^{i}\otimes dx^{i}+\mathring{g}_{a}(x,y)\mathbf{\mathring{e}}%
^{a}\otimes \mathbf{\mathring{e}}^{a}, \\
\mbox{ for } &&\mathbf{\mathring{e}}^{\alpha }=(dx^{i},\mathbf{e}^{a}=dy^{a}+%
\mathring{N}_{i}^{a}(u)dx^{i}), \\
&&\mathbf{\mathring{e}}_{\alpha }=(\mathbf{\mathring{e}}_{i}=\partial
/\partial y^{a}-\mathring{N}_{i}^{b}(u)\partial /\partial y^{b},\ {e}%
_{a}=\partial /\partial y^{a}).
\end{eqnarray*}%

As an explicit example, we  take $\mathbf{\mathring{g}}$ to be a Friedman--Lema\^{\i}%
tre--Robertson--Walker (FLRW) type diagonal metric with $\mathring{N}%
_{i}^{b}=0.$ The target off--diagonal metric is of type (\ref{dm1}) with $%
\mathbf{e}^{a}$ taken as in (\ref{nadif}). With additional parameterizations
via the so-called gravitational "polarization" functions $\eta _{\alpha }=(\eta
_{i},\eta _{a}),$ the metric $\widehat{\mathbf{g}} $ takes the form
\begin{eqnarray}
\widehat{\mathbf{g}} &=&g_{\alpha }(u)\mathbf{e}^{\alpha }\otimes \mathbf{e}%
^{\beta }=g_{i}(x)dx^{i}\otimes dx^{i}+g_{a}(x,y)\mathbf{e}^{a}\otimes
\mathbf{e}^{a}  \label{dm} \\
&=&\eta _{i}(x^{k})\mathring{g}_{i}dx^{i}\otimes dx^{i}+\eta
_{a}(x^{k},y^{b})\mathring{h}_{a}\mathbf{e}^{a}\otimes \mathbf{e}^{a}.
\notag
\end{eqnarray}%
In the special case in which $\eta _{\alpha }\rightarrow 1$ and $N_{i}^{a}=\mathring{N}_{i}^{a},$ we
get a trivial nonholonomic transformation (deformation).

For the data $(\widehat{\mathbf{g}},\widehat{\mathbf{D}})$  the effective source for a scalar field $\phi $ and a gauge field $\mathbf{F}_{\mu
\nu }^{\check a}$ in  modified gravitational interactions (\ref%
{nceq})  is the energy--momentum tensor $^{e}\mathbf{T}%
_{\alpha \beta }$ where
\begin{equation}
\ ^{e}\mathbf{T}_{\alpha \beta }=\frac{1}{2}\left[ \mathbf{e}_{\alpha }\phi
\ \mathbf{e}_{\beta }\phi +\mathbf{e}_{\beta }\phi \ \mathbf{e}_{\alpha
}\phi \ -\widehat{\mathbf{g}}_{\alpha \beta }\widehat{\mathbf{g}}^{\mu \nu }%
\mathbf{e}_{\mu }\phi \ \mathbf{e}_{\nu }\phi +\widehat{\mathbf{g}}_{\alpha
\beta }\ ^{e}V(\phi )\right] +\mathbf{F}_{\ \alpha \nu }^{\check a}\mathbf{F}_{\ \
\beta }^{{\check a} \nu }-\frac{1}{4}\widehat{\mathbf{g}}_{\alpha \beta }\mathbf{F}_{\
\nu \mu }^{\check a}\mathbf{F}_{\ }^{\check a\nu \mu },   \label{effemt}
\end{equation}%
where $\check a$ is a internal group index. This tensor  is constructed with respect to  the  N--adapted (co) frames (\ref{nader}),  (\ref{nadif}) following the same procedure as in Refs. \cite{guend0,guend1,mli,guend2,guend5,guend7},
and $\mathbf{\Upsilon }_{\alpha \beta }=\frac{\kappa }{2}\ ^{e}\mathbf{T}%
_{\alpha \beta }$ where $\kappa$  is the gravitational constant. The explicit coordinate dependence  for $\mathbf{\Upsilon }$ is
\begin{equation}
\mathbf{\Upsilon }_{~1}^{1}=\mathbf{\Upsilon }_{~2}^{2}=\Upsilon (x^{k},t); \,\,\,\,%
\mathbf{\Upsilon }_{~3}^{3}=\mathbf{\Upsilon }_{~4}^{4}=~^{v}\Upsilon
(x^{k})],  \label{source}
\end{equation}%
Elements with $\alpha \ne \beta$ are all taken to be zero.
The effective nonlinear scalar potential $\ ^{e}V$ is determined by two
scalar potentials $V(\phi )$ and $U(\phi )$ as
\begin{equation}
\ ^{e}V=(V+M)^{2}/4U.  \label{effp}
\end{equation}%
where  $M$ is a constant.
The Einstein d--tensor $\widehat{\mathbf{G}}_{\alpha \beta }$ is given in
N--adapted form by formula (\ref{enstdt}).
The resulting nonlinear system of PDEs can
be integrated in explicit form for arbitrary  parameterizations of type
$\mathbf{\Upsilon }_{~\delta }^{\beta }=diag[\mathbf{\Upsilon }_{\alpha }]$.
\footnote{We can consider other distributions which do not allow for the  construction of
solutions in explicit form. Our geometric approach will be applied to
such N--connection splitting and frame/ coordinate transforms that
parameterize the effective sources in some form and will admit the
decoupling of the (modified) Einstein equations.}

As a specific example, we take the TMT  effective action
\begin{equation}
S=\frac{1}{\kappa }\int d^{4}u\sqrt{|\widehat{\mathbf{g}}_{\alpha \beta }|}%
\left[ \widehat{R}+\ ^{m}\widehat{L}\right]   \label{effectconf}
\end{equation}%
studied in \cite{guend0,guend1,guend2,guend5,guend7} for $\ ^{m}\widehat{L}$ resulting in the energy-momentum tensor (\ref{effemt})  and where
  $\widehat{R}$ is the scalar curvature.  Modified Einstein equations are derived in the light of
LC--conditions (\ref{lccond}). The energy--momentum tensor
 follows from variation in N--adapted form using the N--elongated partial derivatives and differentials,
\begin{equation*}
\ ^{e}\mathbf{T}_{\mu \nu }:=-\frac{2}{\sqrt{|\widehat{\mathbf{g}}_{\alpha
\beta }|}}\frac{\delta (\sqrt{|\widehat{\mathbf{g}}_{\alpha \beta }|}\ ^{m}%
\widehat{L})}{\delta \ \widehat{\mathbf{g}}^{\mu \nu }},
\end{equation*}%
We consider a new `scaled' d--metric $\mathbf{g}_{\alpha \beta }$ where
\begin{equation}
\widehat{\mathbf{g}}_{\alpha \beta }=e^{-2\widehat{\sigma }(u)}\mathbf{g}%
_{\alpha \beta },\mbox{
and  }e^{-2\widehat{\sigma }(u)}=2U/(V+M)=\Phi /\sqrt{|\mathbf{g}_{\alpha
\beta }|},  \label{rescalling}
\end{equation}%
where  $e^{-2\widehat{\sigma }}$ is the scale factor determined in terms of the  constant and
potentials used in the effective potential $\ ^{e}V$ \ (\ref{effp}). The
function
\begin{equation*}
\Phi =\varepsilon ^{\mu \nu \alpha \beta }\mathbf{e}_{\mu }\mathbf{A}_{\nu
\alpha \beta }=\varepsilon ^{\mu \nu \alpha \beta }\varepsilon ^{\underline{a%
}\underline{b}\underline{c}\underline{d}}\mathbf{e}_{\mu }\varphi _{%
\underline{a}}\ \mathbf{e}_{\nu }\varphi _{\underline{b}}\ \mathbf{e}%
_{\alpha }\varphi _{\underline{c}}\ \mathbf{e}_{\beta }\varphi _{\underline{d%
}},
\end{equation*}%
with four scalar fields $\varphi _{\underline{a}},$ ($\underline{a}%
=1,2,3,4), $ defines the second measure in TMTs. The
effective gravitational theory (\ref{effectconf}) with the source $\ ^{e}\mathbf{%
T}_{\alpha \beta }$ (\ref{sdcurv}) and re-scaling properties (\ref%
{rescalling}) is equivalent to the theory given by the following action%
\begin{equation}
S=\int \ ^{1}L\Phi d^{4}u+\int \ ^{2}L\sqrt{|\mathbf{g}_{\alpha \beta }|}%
d^{4}x+\int N\phi \varepsilon ^{\mu \nu \alpha \beta }F_{\mu \nu
}^{\check a}F_{\alpha \beta }^{\check a}d^{4}u,  \label{acttmt}
\end{equation}%
where
\begin{eqnarray}
\ ^{1}L &=&-\frac{1}{\kappa }\widehat{R}(\mathbf{g})+\frac{1}{2}\mathbf{g}%
^{\mu \nu }\mathbf{e}_{\mu }\phi \ \mathbf{e}_{\nu }\phi -V(\phi )%
\mbox{
and }  \label{act1} \\
\ ^{2}L &=&U(\phi )-\frac{1}{4}F_{\mu \nu }^{\check a}F^{\check a\mu \nu }.   \notag
\end{eqnarray}%
 In the above, the  $N$ term\footnote{%
Such a "non--boldface" symbol should  not be confused with  the
N--connection $\mathbf{N}=\{N_{I}^{a}\}$; we maintain standard notations in gravity theories with N-connections (boldface symbols). In TMT models  $N$ has a completely different meaning as  introduced in \cite{guend0,guend1,mli,guend2,guend5,guend7}}. It is a CP violating parameter and is determined to be very small from constraints from phenomenology. The
non--Riemannian configuration is determined from the canonical d--connection $%
\widehat{\Gamma }_{\beta \gamma }^{\alpha }$ for $\mathbf{g}_{\alpha \beta}. $

Identifying the scalar indices as interior indices ("overline check") and varying (\ref{acttmt}) with respect to $\varphi _{\check a}$ in N--adapted form, we obtain the equation%
\begin{equation}
\mathbf{A}_{\check a}^{\mu }\ \mathbf{e}_{\mu }\ ^{1}L=0.  \label{ea1}
\end{equation}
The solution of this equation is $\mathbf{e}_{\mu }\ ^{1}L=0,$ or $\
^{1}L=M=const.$ Thus for any $M\neq 0,$ we obtain a spontaneous breaking of
global scale invariance of the theory. This follows from the mismatch between the left hand side  and the right hand side of the equation. If we fix $M$ as an integration constant for the right hand side, the left hand side has a
non--trivial transformation. In terms of the metric $\widehat{\mathbf{g}}%
_{\alpha \beta },$ the equation for the scalar filed becomes
\begin{equation}
\mathbf{e}_{\mu }(\sqrt{|\mathbf{g}_{\alpha \beta }|}\widehat{\mathbf{g}}%
^{\mu \nu }\phi )+\sqrt{|\mathbf{g}_{\alpha \beta }|}\frac{d\ ^{e}V(\phi )}{%
d\phi }+N\varepsilon ^{\mu \nu \alpha \beta }F_{\mu \nu }^{\check a}F_{\alpha \beta
}^{\check a}=0.  \label{eq1a}
\end{equation}%
Not considering effective gauge interactions, i.e. for $N=0,$ we define the
vacuum states for $V+M=0,$ where $\ ^{e}V=0$ and $d\ ^{e}V/d\phi =0$ (it is
also considered that $d\ ^{e}V/d\phi $ is finite and $U\neq 0).$ We conclude
that the basic feature of TMTs do not depend on the type of nonholonomic
distributions  on spacetimes if we work with metric compatible
canonical d--connections or the LC connections. For both cases, we solve
the 'old" cosmological constant problem, implying that the vacuum state
with zero cosmological constant is achieved for different types of linear
connections and without resort to fine tuning. Independently of wether we
change the value of constant $M,$ or add a constant to $V,$ we still satisfy
the conditions $\ ^{e}V=0$ and $d\ ^{e}V/d\phi =0$ if $V+M=0.$ Here we also
note that if we consider  $N\neq 0$, it implies that an external source drives the scalar field
away from such vacuum points and can be addressed in terms of instanton effects.

N--adapted variations with respect to $\mathbf{g}^{\mu \nu }$ result in the equation%
\begin{equation}
\Phi \left[ -\frac{1}{\kappa }\widehat{R}(\mathbf{g})+\frac{1}{4}(\mathbf{e}%
_{\mu }\phi \ \mathbf{e}_{\nu }\phi +\ \mathbf{e}_{\nu }\phi \ \mathbf{e}%
_{\mu }\phi )\right] -\frac{1}{2}\sqrt{|\mathbf{g}_{\alpha \beta }|}U(\phi )%
\mathbf{g}_{\mu \nu }+\sqrt{|g|}\left[ \mathbf{F}_{\mu \alpha }^{\underline{a%
}}\mathbf{F}_{\ \beta }^{\underline{a}\ \alpha }-\frac{1}{4}\mathbf{g}_{\mu
\nu }\mathbf{F}_{\alpha \beta }^{\underline{a}}\mathbf{F}^{\underline{a}%
\alpha \beta }\right] =0,  \label{eq2}
\end{equation}%
where $\underline{a} = \check a$ for this class of TMT theories.
Additional constraints for LC--configurations when the equations (\ref%
{lccond}) for the data $(\mathbf{g},\widehat{\mathbf{D}}[\mathbf{g}])$ are
satisfied transform (\ref{eq2}) into the system (17) in \cite{mli}. A small
vacuum density determined by instantons was analyzed for LC--configurations
of (\ref{eq1a}). It is a cumbersome task to find cosmological solutions of
the system defined by equations (\ref{ea1}) - (\ref{eq2}). Nevertheless, it is possible to
construct generic off--diagonal cosmological solutions for the systems of
modified commutative and noncommutative Einstein -- Yang --Mills - Higgs
fields using the AFDM \cite{vnb15,vnceym,vhep}%
. Our strategy is to find solutions for the theory (\ref{effectconf})
resulting in modified Einstein equations (\ref{nceq}) with effective
stress--energy tensor (\ref{effemt}) and effective source (\ref{source}).
 Metrics such as  $\widehat{\mathbf{g}}_{\alpha \beta }$, in general, transform into
$\mathbf{g}_{\alpha \beta }$ for the theory (\ref{acttmt}) using N--adapted
conformal transforms of type (\ref{rescalling}).


We  integrate in explicit form the equations (\ref{nceq}) with a source (%
\ref{source}) for the N--adapted coefficients of a metric $\widehat{\mathbf{g}%
}$ (\ref{dm}) parameterized in the form
\begin{equation}
g_{i}=e^{\psi {(x^{k})}}, \,\,\,\, g_{a}=\omega (x^{k},y^{b})h_{a}(x^{k},t),\
N_{i}^{3}=n_{i}(x^{k},t), \,\,\,\,  N_{i}^{4}=w_{i}(x^{k},t)  \label{data2}
\end{equation}%
and supplementing  with frame/coordinate transformations that satisfy  the
conditions $h_{a}^{\diamond }\neq 0,\Upsilon _{2,4}\neq 0.$\footnote{%
For simplicity, we shall omit "hats" on coefficients of type $%
g_{i},g_{a},n_{i},$ $w_{i}$ etc related to $\widehat{\mathbf{g}}$ if it
will not lead to ambiguities.} For convenience,  the partial derivatives $\partial
_{\alpha }=\partial /\partial u^{\alpha }$ are labeled as
\begin{equation*}
\partial _{1}s=s^{\bullet }=\partial s/\partial x^{1},\partial
_{2}s=s^{\prime }=\partial s/\partial x^{2},\partial _{3}s=\partial
s/\partial y^{3},\partial _{4}s=\partial s/\partial t=\partial
_{t}s,\partial ^{2}s/\partial t^{2}=\partial _{tt}^{2}s.
\end{equation*}%
The nontrivial components of the Ricci and Einstein d--tensors are computed
using the N--adapted coefficients of the canonical d--connection (\ref%
{candcon}) for the metric ansatz (\ref{dm}) with data (\ref{data2}) for $%
\omega =1$ introduced respectively in (\ref{driccic}), (\ref{sdcurv}) and (%
\ref{enstdt}). Eventually, we arrive at the following system of nonlinear PDEs
\begin{eqnarray}
\widehat{R}_{1}^{1} &=&\widehat{R}_{2}^{2}=\frac{1}{2g_{1}g_{2}}[\frac{%
g_{1}^{\bullet }g_{2}^{\bullet }}{2g_{1}}+\frac{\left( g_{2}^{\bullet
}\right) ^{2}}{2g_{2}}-g_{2}^{\bullet \bullet }+\frac{g_{1}^{\prime
}g_{2}^{\prime }}{2g_{2}}+\frac{(g_{1}^{\prime })^{2}}{2g_{1}}-g_{1}^{\prime
\prime }]=-\ ^{v}\Upsilon ,  \label{eq1b} \\
\widehat{R}_{3}^{3} &=&\widehat{R}_{4}^{4}=\frac{1}{2h_{3}h_{4}}[\frac{%
\left( \partial _{t}h_{3}\right) ^{2}}{2h_{3}}+\frac{\partial _{t}h_{3}\
\partial _{t}h_{4}}{2h_{4}}-\partial _{tt}^{2}h_{3}]=-\Upsilon  \label{eq2b}
\\
\widehat{R}_{3k} &=&\frac{h_{3}}{2h_{4}}\partial _{tt}^{2}n_{k}+(\frac{h_{3}%
}{h_{4}}\partial _{t}h_{4}-\frac{3}{2}\partial _{t}h_{3})\frac{\partial
_{t}n_{k}}{2h_{4}}=0,  \label{eq3b} \\
\widehat{R}_{4k} &=&\frac{w_{k}}{2h_{3}}[\partial _{tt}^{2}h_{3}-\frac{%
\left( \partial _{t}h_{3}\right) ^{2}}{2h_{3}}-\frac{\partial _{t}h_{3}\
\partial _{t}h_{4}}{2h_{4}}]+\frac{\partial _{t}h_{3}}{4h_{3}}(\frac{%
\partial _{k}h_{3}}{h_{3}}+\frac{\partial _{k}h_{4}}{h_{4}})-\frac{\partial
_{k}\partial _{t}h_{3}}{2h_{3}}=0.  \label{eq4b}
\end{eqnarray}%
The torsionless (Levi--Civita, LC) conditions (\ref{lc}), (\ref%
{lccond}), transform into
\begin{eqnarray}
\partial _{t}w_{i} &=&(\partial _{i}-w_{i}\partial _{t})\ln \sqrt{|h_{4}|}%
,(\partial _{i}-w_{i}\partial _{t})\ln \sqrt{|h_{3}|}=0,  \label{lccondb} \\
\partial _{k}w_{i} &=&\partial _{i}w_{k},\partial _{t}n_{i}=0,\partial
_{i}n_{k}=\partial _{k}n_{i}.  \notag
\end{eqnarray}

The system of nonlinear PDE (\ref{eq1b})--(\ref{eq4b}) posses an important
decoupling property which admits step by step integration of such
equations.  To achieve this, first we introduce the coefficients
\begin{equation}
\alpha _{i}=(\partial _{t}h_{3})\ (\partial _{i}\varpi),\ \beta =(\partial
_{t} h_{3})\ (\partial _{t}\varpi),\ \gamma =\partial _{t}\left( \ln
|h_{3}|^{3/2}/|h_{4}|\right) ,  \label{abc}
\end{equation}%
where
\begin{equation}
\varpi {=\ln |\partial _{t}h_{3}/\sqrt{|h_{3}h_{4}|}|} . \label{genf}
\end{equation}%
The coefficients serve as  generating functions. For $\partial _{t}h_{a}\neq 0$ and $%
\partial _{t}\varpi \neq 0,$\footnote{%
Nontrivial solutions result if such conditions are not satisfied; in such
cases, we need to consider other special methods for generating solutions.} we
 rewrite the equations in the form
\begin{eqnarray}
\psi ^{\bullet \bullet }+\psi ^{\prime \prime } &=&2~^{v}\Upsilon  \label{e1}
\\
\partial _{t}\varpi \ \partial _{t}h_{3} &=&2h_{3}h_{4}\Upsilon  \label{e2}
\\
\partial _{tt}^{2}n_{i}+\gamma \partial _{t}n_{i} &=&0,  \label{e3} \\
\beta w_{i}-\alpha _{i} &=&0,  \label{e4} \\
\partial _{i}\omega -n_{i}\partial _{3}\omega -(\partial _{i}\varpi
/\partial _{t}\varpi )\partial _{t}\omega &=&0.  \label{confeq}
\end{eqnarray}%
The function $\psi (x^{k})$ are found by solving  a two dimensional Poisson
equation (\ref{e1}) for any prescribed source $\ ^{v}\Upsilon (x^{k}).$ The
equations (\ref{genf}) and (\ref{e2}) convert any two functions
to two others from a set of four, $h_{a},\varpi $ and $\Upsilon .$
In one explicit form,  $h_{3}$ and $h_{4}$ are determined for any prescribed $\varpi
(x^{k},t)$ and $\Upsilon (x^{k},t).$ Once $h_{a}$ are determined, we integrate twice
 w.r.t $t$ in (\ref{e3}) and find $n_{i}(x^{k},t)$. In the final step  we solve for $w_{i}(x^{i},y^{a})$
 by solving a system of linear algebraic equations (\ref{e4}).
The equation (\ref{confeq}) is necessary to accommodate a nontrivial
conformal (in the vertical "subspace") factor $\omega (x^{i},y^{a})$ that
depends on all four coordinates. For convenience, we shall use $\Psi :=e^{\varpi }$ as our
re--defined generating function.

We conclude  section \ref{s2} with the following remarks. We have shown that TMT
theories as determined by actions of type (\ref{acttmt}) can be formulated in
nonholonomic variables as effective EYMH systems with modified Einstein
field equations (\ref{nceq}). This allows one to apply the AFDM and decouple
such systems of nonlinear PDEs in very general form and write them
equivalently as systems of type  (\ref{e1})--(\ref{confeq}). This procedure and the resulting equations provide important results for mathematical cosmology. For instance, by considering  the coordinate $y^{4}=t
$ to be  time like, one can show that TMT theories and other modified gravity
models can be integrated in general forms.

\section{Off--Diagonal Cosmological Solutions with Small Vacuum Density}
\label{s3}

In this section we provide a series of examples of new classes of exact solutions of modified Einstein equations with (non) homogeneous cosmological configurations constructed by applying the AFDM. We emphasize that all solutions generated in this section will be for a TMT theory with sources (\ref{effemt})  parameterized in the form (\ref{source}), when the effective nonlinear scalar potential is taken in the form (\ref{effp}).  In a similar form, we can construct solutions with effective  sources for other types of modified gravity theories like in \cite{elizaldev,veffmassiv1}.

For any  $\partial _{t}\varpi \neq 0,\partial _{t}h_{a}\neq 0$ and $\Upsilon \neq 0,$ we write (\ref{e2}) and (\ref{genf}) as
\begin{equation}
\ h_{3}h_{4}=(\partial _{t}\varpi )(\partial _{t}h_{3})/2\Upsilon
\mbox{
and }|h_{3}h_{4}|=(\partial _{t}h_{3})^{2}e^{-2\varpi }.  \label{eq4bb}
\end{equation}%
Using $\Psi :=e^{\varpi }$ and introducing the first equation into the second  in (\ref{eq4bb}), we obtain the relation $|\partial _{t}h_{3}|=\partial _{t}[\Psi ^{2}]/4|\Upsilon |.$ Integrating with respect  to $t,$ we get
\begin{equation}
h_{3}[\Psi ,\Upsilon ]=\ ^{0}h_{3}(x^{k})-\frac{1}{4}\int dt\frac{\partial
_{t}(\Psi ^{2})}{\Upsilon },  \label{h3aux}
\end{equation}%
where $\ ^{0}h_{3}=\ ^{0}h_{3}(x^{k})$ is an integration function. We  use
the first equation in (\ref{eq4bb}) and compute
\begin{equation}
h_{4}[\Psi ,\Upsilon ]=\frac{1}{2\ \Upsilon }\frac{\partial _{t}\Psi }{\Psi }%
\frac{\partial _{t}h_{3}}{h_{3}}.  \label{h4aux}
\end{equation}%
Formulae for $h_{a}$ are  expressed in a more convenient form by considering an effective cosmological constant $\Lambda _{0}=const\neq 0$ and a re--defined generating function, $\Psi \rightarrow \tilde{\Psi},$ subject
to the condition
\begin{equation*}
\frac{\partial _{t}[\Psi ^{2}]}{\Upsilon }=\frac{\partial _{t}[\tilde{\Psi}%
^{2}]}{\Lambda _{0}},
\end{equation*}%
where the integration function$\ ^{0}h_{3}(x^{k})$ from (\ref{h3aux}) is formally introduced either in $\widetilde{\Psi }$ or equivalently  in $\Upsilon .$

Our final results are
\begin{equation}
h_{3}[\widetilde{\Psi },\Lambda _{0}]=\frac{\widetilde{\Psi }^{2}}{4\Lambda
_{0}}\mbox{ and }
h_{4}[\widetilde{\Psi },\Lambda _{0},\Xi ]=\frac{(\partial _{t}\widetilde{%
\Psi })^{2}}{\Xi }  \label{solha}
\end{equation}%
and hold for an effective cosmological constant $\Lambda _{0}\neq 0$ so that
re--definition of  the generating functions, $\Psi \longleftrightarrow \widetilde{%
\Psi },$  are unambiguous
 where
\begin{equation}
\Psi ^{2}=\Lambda _{0}^{-1}\int dt\Upsilon \partial _{t}(\widetilde{\Psi }%
^{2})\mbox{
and }\widetilde{\Psi }^{2}=\Lambda _{0}\int dt\Upsilon ^{-1}\partial
_{t}(\Psi ^{2}).  \label{rescgf}
\end{equation}%
The functional
\begin{equation*}
\Xi \lbrack \Upsilon ,\widetilde{\Psi }]=\int dt\Upsilon \partial _{t}(%
\widetilde{\Psi }^{2})
\end{equation*}%
in the formula for $h_{4}$ in (\ref{solha}) is interpreted as a
re--defined source $\ \Upsilon \rightarrow \Xi $ for a prescribed
generating function $\widetilde{\Psi }$ when $\Upsilon =\partial _{t}\Xi
/\partial _{t}(\widetilde{\Psi }^{2}).$ Such effective sources contain
information on effective matter field contributions in modified gravity
theories. We work  with the  generating quantities, $%
(\Psi ,\ ^{v}\Lambda )$ and $[\widetilde{\Psi },\Lambda _{0},\Xi ]$ related
via formulae (\ref{rescgf}) in terms of the  prescribed effective cosmological constant $%
\Lambda _{0}$. The numerical value of $\Lambda _{0}$ is fixed to meet present day constraints from cosmology.

Using formulae $h_{a}$ (\ref{solha}), we compute the coefficients $%
\alpha _{i},\beta $ and $\gamma $ from (\ref{abc}). This allows us to find
solutions to equations (\ref{e3}) by integrating two times with respect to $t,$ and (\ref%
{e4}), solving a system of  linear  algebraic equations for $w_{i}.$ As a
result, the N--coefficients are expressed recurrently as functionals
 (an example of which is  $[\widetilde{\Psi },\Lambda _{0},\Xi ]$) and are as follows,
\begin{eqnarray}
n_{k} &=&\ _{1}n_{k}+\ _{2}n_{k}\int dth_{4}/(\sqrt{|h_{3}|})^{3}=\
_{1}n_{k}+\ _{2}\widetilde{n}_{k}\int dt(\partial _{t}\widetilde{\Psi })^{2}/%
\widetilde{\Psi }^{3}\Xi ,\mbox{ and }  \notag \\
w_{i} &=&\partial _{i}\varpi /\partial _{t}\varpi =\partial _{i}\Psi
/\partial _{t}\Psi =\partial _{i}\Psi ^{2}/\partial _{t}\Psi ^{2}=\int
dt\partial _{i}[\Upsilon \partial _{t}(\widetilde{\Psi }^{2})]/\Upsilon
\partial _{t}(\widetilde{\Psi }^{2})=\partial _{i}\Xi /\partial _{t}\Xi ,
\label{solhn}
\end{eqnarray}%
where $\ _{1}n_{k}(x^{i})$ and $\ _{2}n_{k}(x^{i}),$ or $_{2}\widetilde{n}%
_{k}(x^{i}),$ are integration functions with possible re--definitions by
coordinate transforms.

After a  tedious calculation for $g_{a}=\omega ^{2}(x^{k},y^{a})h_{a}$ that involves  the vertical
conformal factor $\omega (u^{\alpha })$ depending on all spacetime
coordinates, the vertical metric $h_{a}$ (\ref{solha}) and the N--coefficients $%
N_{i}^{a}$ (\ref{solhn}) reveals the fact  that the formulae for the Ricci d--tensor $%
\widehat{\mathbf{R}}_{\alpha \beta }$ (\ref{driccic}) are invariant if the
 first order PDE (\ref{confeq}) are satisfied. For  nontrivial $\omega ,$
the solutions to the modified gravitational equations, (\ref{nceq})
parameterized as a d--metric (\ref{dm}), do not posses in general any
Killing symmetries and contain dependencies of  $\omega $ on  $[\psi
,h_{a},n_{i},w_{i}]$ with as many as six independent variables for $%
\mathbf{g}_{\alpha \beta }.$

Putting together the solutions for the 2--d Poisson equation (\ref{e1}) and the
formulae for the coefficients (\ref{solha}), (\ref{solhn}) we conclude as our final result  that
the system of nonlinear PDEs (\ref{e1})-- (\ref{e4}) for non--vacuum 4--d
configurations for the data $(\mathbf{g,N,}\widehat{\mathbf{D}}),$ and with Killing
symmetry on $\partial _{3}$ when  $\omega =1,$ integrates to the line element
\begin{eqnarray}
ds^{2} &=&g_{\alpha \beta }(x^{k},t)du^{\alpha }du^{\beta }=e^{\psi
(x^{k})}[(dx^{1})^{2}+(dx^{2})^{2}]+  \label{nkillingsol} \\
&&\omega ^{2}\frac{\widetilde{\Psi }^{2}}{4\Lambda _{0}}[dy^{3}+\left( \
_{1}n_{k}+_{2}\widetilde{n}_{k}\int dt\frac{(\partial _{t}\widetilde{\Psi }%
)^{2}}{\widetilde{\Psi }^{3}\ \Xi }\right) dx^{k}]^{2}+\omega ^{2}\frac{%
(\partial _{t}\widetilde{\Psi })^{2}}{\Xi }\ [dt+\frac{\partial _{i}\Xi }{%
\partial _{t}\Xi }dx^{i}]^{2}.  \notag
\end{eqnarray}%
Such inhomogeneous cosmological solutions with nonholonomically induced
torsion are determined by  $ \psi (x^{k}),$ $\widetilde{\Psi }(x^{k},t),$ $ \omega (x^{k},y^{3},t),$
$\Xi (x^{k},t)$  that depend on the effective cosmological constant $\Lambda _{0}$ and
integration functions $\ _{1}n_{k}$, $\ _{2}\widetilde{n}_{k}$ . Straightforward  computations reveal that,  in general, the nonholonomy
coefficients $W_{\alpha \beta }^{\gamma }$ (\ref{anhcoef}) are non vanishing.
Therefore the class of solutions (\ref{nkillingsol}) can not be diagonalized in
N--adapted form unless supplemented with  additional assumptions  on
generating/ integration functions and constants. The nontrivial coefficients
of the canonical d--torsion(\ref{dt}) are also non vanishing. They are determined by
introducing the coefficients of the d--metric into N--adapted formulas (\ref%
{candcon}) and then into $\widehat{\mathbf{T}}_{\alpha \beta }^{\gamma }$ (%
\ref{dtors}).

\label{sslccond}Let us prove that the zero d--torsion conditions (\ref%
{lccondb}) for LC--configurations can be solved in explicit form by imposing
additional constraints on d--metrics (\ref{nkillingsol}). For the $n$%
--coefficients, such conditions are satisfied if $\ _{2}n_{k}(x^{i})=0$ and $%
\partial _{i}\ _{1}n_{j}(x^{k})=\partial _{j}\ _{1}n_{i}(x^{k}).$ In
N--adapted form, such coefficients do not depend on generating functions and
sources but only on a corresponding class of integration functions, e.g.,  $\ _{1}n_{j}(x^{k})=$ $\partial _{i}n(x^{k}),$ for any $n(x^{k}).$ It is a more difficult task to find explicit solutions for the
LC--conditions (\ref{lccondb}) involving variables $w_{i}(x^{k}).$ Such
nonholonomic constraints can not be solved in explicit form for arbitrary
data $(\Psi ,\Upsilon ),$ or arbitrary $(\tilde{\Psi},\Xi ,\Lambda _{0}).$
We first use the property that $\mathbf{e}_{i}\Psi =(\partial
_{i}-w_{i}\partial _{t})\Psi \equiv 0$ for any $\Psi $ if $w_{i}=\partial
_{i}\Psi /\partial _{t}\Psi $ (it follows from formulas (\ref{solhn})). This results in the expression
\begin{equation*}
\mathbf{e}_{i}H=(\partial _{i}-w_{i}\partial _{t})H=\frac{\partial H}{%
\partial \Psi }(\partial _{i}-w_{i}\partial _{t})\Psi \equiv 0
\end{equation*}%
for any functional $H[\Psi ].$ The second step is to restrict our
construction to a subclass of variables when $H=\tilde{\Psi}[\Psi ]$
is a functional which allows us to generate LC--configurations in explicit
form. By taking $h_{3}[\tilde{\Psi}]=\tilde{\Psi}^{2}/4\Lambda _{0}$ (\ref%
{solha}) as a necessary type of functional $H=$ $\tilde{\Psi}=\ln \sqrt{|\
h_{3}|},$ we satisfy the condition $\mathbf{e}_{i}\ln \sqrt{|\ h_{3}|}=0$ in
(\ref{lccondb}).

Next, we solve for the constraint on $h_{4}.$ The derivative $%
\partial _{4}$ of $\ w_{i}=\partial _{i}\Psi /\partial _{t}\Psi $ (\ref%
{solhn}) results in
\begin{equation*}
\partial _{t}w_{i}=\frac{(\partial _{t}\partial _{i}\Psi )(\partial _{t}\Psi
)-(\partial _{i}\Psi )\partial _{t}^{2}\Psi }{(\partial _{t}\Psi )^{2}}=%
\frac{\partial _{t}\partial _{i}\Psi }{\partial _{t}\Psi }-\frac{\partial
_{i}\Psi }{\partial _{t}\Psi }\frac{\partial _{t}^{2}\Psi }{\partial
_{t}\Psi }.
\end{equation*}%
Substituting in this formula the  generating function $\Psi =\check{\Psi}$ gives
\begin{equation}
\partial _{t}\partial _{i}\check{\Psi}=\partial _{i}\partial _{t}\check{\Psi}%
,  \label{explcond}
\end{equation}%
and we deduce that  $\partial _{t}w_{i}=\mathbf{e}_{i}\ln |\partial _{t}\check{\Psi}%
|. $ By extracting $h_{4}[\check{\Psi},\ ^{v}\Lambda ]$ from (\ref{h4aux}) with $%
\check{\Psi},$ we  arrive at
\begin{equation*}
\mathbf{e}_{i}\ln \sqrt{|\ h_{4}|}=\mathbf{e}_{i}[\ln |\partial _{t}\check{%
\Psi}|-\ln \sqrt{|\Upsilon |}],
\end{equation*}%
In order to prove this formula we  have used  (\ref{explcond}) and $\mathbf{e}%
_{i}\check{\Psi}=0.$ From the last two formulae, we obtain $\partial _{t}w_{i}=%
\mathbf{e}_{i}\ln \sqrt{|\ h_{4}|}$ if
\begin{equation*}
\mathbf{e}_{i}\ln \sqrt{|\Upsilon |}=0.
\end{equation*}%
This is possible for either $\Upsilon =const,$ or if $\Upsilon $ can be expressed
as a functional $\Upsilon (x^{i},t)=\ \check{\Upsilon}[\check{\Psi}].$ If
such conditions are not satisfied, we can re--scale the generating function $%
\check{\Psi}\longleftrightarrow \widetilde{\Psi },$ where
\begin{equation*}
\check{\Psi}^{2}=\Lambda _{0}^{-1}\int dt\check{\Upsilon}\partial _{t}(%
\widehat{\Psi }^{2})\mbox{
and }\widehat{\Psi }^{2}=\Lambda _{0}\int dt\check{\Upsilon}^{-1}\partial
_{t}(\check{\Psi}]^{2}),
\end{equation*}%
when
\begin{equation}
\partial _{t}\partial _{i}\widehat{\Psi }=\partial _{i}\partial _{t}\widehat{%
\Psi }.  \label{explconda}
\end{equation}%
We consider a functional
\begin{equation*}
\widehat{\Xi }[\ \check{\Upsilon},\widehat{\Psi }]=\int dt\ \check{\Upsilon}%
\partial _{t}(\widetilde{\Psi }^{2})
\end{equation*}%
in the formula for $h_{4}$ (\ref{solha}) (as a re--defined source,$\ \check{%
\Upsilon}\rightarrow \widehat{\Xi }),$ for a prescribed generating function $%
\widehat{\Psi },$ when $\check{\Upsilon}=\partial _{t}\widehat{\Xi }%
/\partial _{t}(\check{\Psi}^{2})$ for any effective cosmological constant $%
\Lambda _{0}$ in order to satisfy such conditions.

If we introduce a function $\check{A}=\check{A}(x^{k},t)$ for which
\begin{equation*}
w_{i}=\check{w}_{i}=\partial _{i}\check{\Psi}/\partial _{t}\check{\Psi}%
=\partial _{i}\widehat{\Xi }/\partial _{t}\widehat{\Xi }=\partial _{i}\check{%
A},
\end{equation*}%
then $\partial _{i}w_{j}=\partial _{j}w_{i}$ in (\ref{lccondb}).

Summarizing the results, we conclude that the linear quadratic line  element
\begin{eqnarray}
ds^{2} &=&g_{\alpha \beta }(x^{k},t)du^{\alpha }du^{\beta }  \label{lcsolut}
\\
&=&e^{\psi (x^{k})}[(dx^{1})^{2}+(dx^{2})^{2}]+\omega ^{2}\frac{\widehat{%
\Psi }^{2}}{4\Lambda _{0}}[dy^{3}+\partial _{i}n(x^{k})dx^{i}]^{2}+\omega
^{2}\frac{(\partial _{t}\widehat{\Psi })^{2}}{\widehat{\Xi }}\ [dt+\partial
_{i}\check{A}\ dx^{i}]^{2},  \notag
\end{eqnarray}%
where $\omega $ is a solution of
\begin{equation*}
\partial _{i}\omega -\partial _{i}n\ \partial _{3}\omega -(\partial _{i}%
\widehat{\Xi }/\partial _{t}\widehat{\Xi })\ \partial _{t}\omega =0
\end{equation*}%
and defines generic off--diagonal cosmological solutions with zero
nonholonomically induced torsion. Such inhomogeneous cosmological solutions
are determined by  the generating functions and effective sources $
\psi (x^{k}),$ $\widehat{\Psi }(x^{k},t),$ $\omega (x^{k},y^{3},t)$,
 $\widehat{\Xi }(x^{k},t),$ the parameter $\Lambda _{0},$ and the
integration functions $\ _{1}n_{i}=\partial _{i}n(x^{k})$ respectively. The main result of this
section is the demonstration that TMT theories admit generic
off-diagonal cosmological solutions of type (\ref{nkillingsol}), with
nontrivial nonholonomically induced torson, or of type (\ref{lcsolut}), for
LC-configurations. Another fundamental physical result is the emergence of
a  nonlinear symmetry for generating functions, see formula (\ref{rescgf}), for cosmological solutions of such nonlinear systems which allows to transform arbitrary effective and matter fields sources into an effective
cosmological  constant $\Lambda _{0}$ treated  as an integration parameter. The value of the integration parameter  can be fixed by getting compatibility with observational cosmological data.

\section{Time like parameterized off--diagonal cosmological solutions}
\label{s4}

In this section we consider a subclass of solutions  pertaining to $g_{\alpha \beta }(x^{k},y^{3},t)$  extracted from either (\ref{nkillingsol}), or (\ref{lcsolut}) which, via frame transformations  $%
g_{\alpha \beta }(u)=e_{\ \alpha }^{\alpha ^{\prime }}(u)e_{\ \beta }^{\beta
^{\prime }}(u)g_{\alpha ^{\prime }\beta ^{\prime }}(t),$ result in metrics $%
g_{\alpha ^{\prime }\beta ^{\prime }}(t)$ that depend only on time like
coordinate $t$. For applications in modern cosmology,
we  consider $g_{\alpha ^{\prime }\beta ^{\prime }}(t)$ as certain
off--diagonal deformations of the FLRW, or the Bianchi type Universes \cite{vcosms,elizaldev}. In explicit form, we  construct physical models with $\mathbf{\acute{g}}=\{g_{\alpha ^{\prime }\beta
^{\prime }}(t)\}\rightarrow $ $\mathbf{\mathring{g}}=\{\mathring{g}_{i},%
\mathring{h}_{a}\}$ for $\eta _{\alpha }\rightarrow 1$ and $\mathbf{e}%
^{\alpha }\rightarrow du^{\alpha }=(dx^{i},dy^{a})$ in (\ref{dm}). The strategy is first to
construct solutions for a class of generating functions and sources with
 spacetime dependent coordinates and then to restrict the integral varieties
to configurations with dependencies only on the  time like coordinate. This procedure requires that
 $\widetilde{\Psi }(x^{k},t)\rightarrow \widetilde{\acute{\Psi}}(t), \widehat{\Psi }(x^{k},t)\rightarrow \widehat{\acute{\Psi}}(t);$ $\Upsilon (x^{k},t)\rightarrow \acute{\Upsilon}(t)$ with
 $\Xi \lbrack \Upsilon ,\widetilde{\Psi }]=\int dt\Upsilon \partial _{t}(\widetilde{\Psi }^{2})\rightarrow \acute{\Xi}(t)=\acute{\Xi}[\acute{\Upsilon} (t),\widetilde{\acute{\Psi}}(t)]  ;  \widehat{\Xi }[\Upsilon ,%
\widehat{\Psi }]=\int dt\Upsilon \partial _{t}(\widehat{\Psi } ^{2})\rightarrow \widehat{\acute{\Xi}}(t)=\widehat{\acute{\Xi}}[\acute{\Upsilon}(t),\widehat{\acute{\Psi}}(t)];$
$\partial _{i}\acute{\Xi}\rightarrow \acute{\digamma}_{i}(t),$   $%
\partial _{i}\widehat{\acute{\Xi}}\rightarrow \widehat{\acute{\digamma}}%
$ and with $ \omega \rightarrow 1. $ The
integration functions $\ _{1}n_{k}(x^{i})$ and $\ _{2}\widetilde{n}%
_{k}(x^{i})$ are considered to be constants of integration, implying  $\partial
_{i}n(x^{k})\rightarrow const.  \mbox{ and } \ \partial _{i}\check{A}(x^{k},t)\rightarrow
\check{\digamma}_{i}(t).$

\subsection{Cosmological solutions for the effective EYMH systems and TMT}

The effective gravitational theory (\ref{effectconf}) with source $\ ^{e}%
\mathbf{T}_{\alpha \beta }$ (\ref{sdcurv}) in TMTs describes a nonlinear
parametrical interacting EYMH system where we interpret $\phi $ as a Higgs
field that can carry internal indices and acquire vacuum expectation $\phi _{\lbrack 0]},$ \
and couple to the gauge field $\mathbf{A}=\mathbf{A}_{\mu }e^{\mu }$ with values in
non--Abelian Lie algebra. On the premises defined by the nonholonomic $%
\mathbf{V},$  the d--operator $\widehat{\mathbf{D}}_{\mu }$ is
elongated additionally to accommodate the gauge potentials in the form $\widehat{\mathcal{D}}%
_{\mu }=\widehat{\mathbf{D}}_{\mu }+ie[\mathbf{A}_{\mu },],$ where the
commutator $[.,.]$ signifies the  non--Abelian structure. The gauge coupling is $e$  and $i^{2}=-1.$ The gauge field $\mathbf{A}_{\mu }$ enters the  covariant derivative $D_{\mu }=\mathbf{e}_{\mu }$ $+ie[\mathbf{A}%
_{\mu },]$ and the  "curvature"
\begin{equation}
\mathbf{F}_{\beta \mu }=\mathbf{e}_{\beta }\mathbf{A}_{\mu }-\mathbf{e}_{\mu }\mathbf{%
A}_{\beta }+ie[\mathbf{A}_{\beta },\mathbf{A}_{\mu }],  \label{gaugestr}
\end{equation}%
where the boldface $\mathbf{F}_{\beta \mu }$ is used for N--adapted
constructions.\footnote{For standard gauge field models but on nonholonomic manifolds we can follow
a variational principle for a gravitating non--Abelian SU(2) gauge field $%
\mathbf{A}=\mathbf{A}_{\mu }\mathbf{e}^{\mu }$ coupled to a triplet Higgs
field $\phi .$ In such cases, the value $\phi _{\lbrack 0]}$ is the vacuum
expectation of the Higgs field which determines the mass $\ ^{H}M=\sqrt{%
\lambda }\eta ,$ when $\lambda $ is the \ constant of scalar field
self--interaction with potential $\mathcal{V}(\phi )=\frac{1}{4}\lambda
Tr(\phi _{\lbrack 0]}^{2}-\phi ^{2})^{2},$ where the trace $Tr$ is taken on
internal indices. In EYMH theory, the gravitational constant $G,\kappa
=16\pi G,$ defines the Plank mass $M_{Pl}=1/\sqrt{G}$ and it is also the
mass of gauge boson, $\ ^{W}M=ev.$ In the literature, various versions of modified gravity and
TMTs are elaborated upon with different types of nonlinear scalar and gauge
fields.}

With respect to N--adapted frames the nonholonomic EYMH equations,
postulated either by following geometric principles, or "derived" following an
N--adapted variational calculus from (\ref{effectconf}), are the following,%
\begin{eqnarray}
\widehat{\mathbf{R}}_{\alpha \beta }-\frac{1}{2}\widehat{\mathbf{g}}_{\alpha
\beta }\ \widehat{R} &=&\frac{\kappa }{2}\left( \ ^{\phi }T_{\beta \delta
}+\ ^{F}T_{\beta \delta }\right) ,  \label{ym1} \\
(\sqrt{|\widehat{\mathbf{g}}|})^{-1}\ D_{\mu }(\sqrt{|\widehat{\mathbf{g}}|}%
F^{\mu \nu }) &=&\frac{1}{2}ie[\phi ,D^{\nu }\phi ],  \label{heq2} \\
(\sqrt{|\widehat{\mathbf{g}}|})^{-1}\ D_{\mu }(\sqrt{|\widehat{\mathbf{g}}|}%
\phi ) &=&\lambda (\ \phi _{\lbrack 0]}^{2}-\phi ^{2})\phi ,  \label{heq3}
\end{eqnarray}%
where the source (\ref{source}) is determined by the stress--energy tensor
\begin{eqnarray}
&&\ ^{\phi }T_{\beta \delta }=Tr[\frac{1}{4}\ (D_{\delta }\phi \ D_{\beta
}\phi +D_{\beta }\phi \ D_{\delta }\phi )-\frac{1}{4}\widehat{\mathbf{g}}%
_{\beta \delta }D_{\alpha }\phi \ D^{\alpha }\phi ]-\widehat{\mathbf{g}}%
_{\beta \delta }\ ^{e}V(\phi ),  \label{source1} \\
&&\ ^{F}T_{\beta \delta }=2Tr\left( \widehat{\mathbf{g}}^{\mu \nu }\mathbf{F}%
_{\beta \mu }\mathbf{F}_{\delta \nu }-\frac{1}{4}\widehat{\mathbf{g}}_{\beta
\delta }\mathbf{F}_{\mu \nu }\mathbf{F}^{\mu \nu }\right) .  \label{source2}
\end{eqnarray}%
The nonlinear potential $\ ^{e}V(\phi )$ is  as in  (\ref{source1}) for a TMT if
it is taken in the form (\ref{effp}).

The system of nonlinear PDEs (\ref{ym1})--(\ref{heq3}) posses a similar
decoupling property as in (\ref{nceq}) if plausible  assumptions are made
for gravitational and matter field interactions. To see this and construct
new classes of modified EYMH equations we take the \textquotedblright
prime\textquotedblright\ solution to be given by data for a diagonal d--metric $%
\ ^{\circ }\mathbf{g=}[\ ^{\circ }g_{i}(x^{1}),\ ^{\circ }h_{a}(x^{k}),$ $\
^{\circ }N_{i}^{a}=0]$ with matter fields $\ ^{\circ }A_{\mu }(x^{1})$ and $%
^{\circ }\Phi (x^{1}).$ For $SU(2)$ gauge field configurations, the diagonal
ansatz for generating solutions can be written in the form%
\begin{eqnarray}
\ ^{\circ }\mathbf{g} &=&\ ^{\circ }g_{i}(x^{1})dx^{i}\otimes dx^{i}+\
^{\circ }h_{a}(x^{1},x^{2})dy^{a}\otimes dy^{a}=  \label{ansatz1} \\
&=&q^{-1}(r)dr\otimes dr+r^{2}d\theta \otimes d\theta +r^{2}\sin ^{2}\theta
d\varphi \otimes d\varphi -\sigma ^{2}(r)q(r)dt\otimes dt,  \notag
\end{eqnarray}%
where the coordinates and metric coefficients are parameterized
respectively as $u^{\alpha }=(x^{1}=r,x^{2}=\theta ,y^{3}=\varphi ,y^{4}=t)$
and $\ ^{\circ }g_{1}=q^{-1}(r),\ ^{\circ }g_{2}=r^{2},\ ^{\circ
}h_{3}=r^{2}\sin ^{2}\theta ,\ ^{\circ }h_{4}=-\sigma ^{2}(r)q(r),$ for $%
q(r)=1-$ $2m(r)/r-\Lambda r^{2}/3,$ and $\Lambda $ is a cosmological
constant. The function $m(r)$ is  interpreted as the total mass
within the radius $r$  for which $m(r)=0$ defines an empty de Sitter
space written in a static coordinate system with a cosmological horizon at $%
r=r_{c}=\sqrt{3/\Lambda }.$ The solution of (\ref{ym1}) associated to the
quadratic metric line element (\ref{ansatz1}) is defined by a single magnetic
potential $\omega (r),$
\begin{equation}
\ ^{\circ }A=\ ^{\circ }A_{2}dx^{2}+\ ^{\circ }A_{3}dy^{3}=\frac{1}{2e}\left[
\omega (r)\tau _{1}d\theta +(\cos \theta \ \tau _{3}+\omega (r)\tau _{2}\sin
\theta )\ d\varphi \right] ,  \label{ans1a}
\end{equation}%
where $\tau _{1},\tau _{2},\tau _{3}$ are Pauli matrices. The corresponding
solution of (\ref{heq3}) is given by
\begin{equation}
\Phi =\ ^{\circ }\Phi =\varpi (r)\tau _{3}.  \label{ans1b}
\end{equation}%
Explicit values for the functions $\sigma (r),q(r),\omega (r),\varpi (r)$
have been found in Ref. \cite{br3} for ansatz (\ref{ansatz1}), (\ref{ans1a}%
) and (\ref{ans1b}) when $\left[ \ ^{\circ }\mathbf{g}(r),\ ^{\circ }A(r),\
\ ^{\circ }\Phi (r)\right] $ define physical solutions with diagonal metrics
depending only on the radial coordinate.  A typical example is the well
known diagonal Schwarz\-schild--de Sitter solution  (\ref{ym1})--(\ref%
{heq3})  that is given by%
\begin{equation*}
\omega (r)=\pm 1,\sigma (r)=1,\phi (r)=0,q(r)=1-2M/r-\Lambda r^{2}/3
\end{equation*}%
and  defines a black hole configuration inside a cosmological horizon
because $q(r)=0$ has two positive solutions and $M<1/3\sqrt{\Lambda }.$

The conditions for nonholonomic deformations of (\ref{ansatz1}) are as follows.
  \ The \textquotedblright target\textquotedblright\ d--metric $\ ^{\eta }\mathbf{g}$ 
  with nontrivial N--coefficients, for $\ ^{\circ }\mathbf{g\rightarrow }\widehat{\mathbf{g}}$ 
  is parameterized as in  (\ref{dm}). The gauge fields are deformed as
\begin{equation}
A_{\mu }(x^{i},y^{3})=\ ^{\circ }A_{\mu }(x^{1})+\ ^{\eta }A_{\mu
}(x^{i},y^{a}),  \label{ans2a}
\end{equation}%
where $\ ^{\circ }A_{\mu }(x^{1})$ is of the type (\ref{ans1a}) and $\ ^{\eta
}A_{\mu }(x^{i},y^{a})$ are  functions for which
\begin{equation}
\mathbf{F}_{\beta \mu }=\ ^{\circ }F_{\beta \mu }(x^{1})+\ \ ^{\eta }\mathbf{%
F}_{\beta \mu }(x^{i},y^{a})=s\sqrt{|g|}\varepsilon _{\beta \mu },
\label{gaugstr1}
\end{equation}%
where $s$ is a constant and $\varepsilon _{\beta\mu }$ is the absolute antisymmetric tensor. The gauge field curvatures $F_{\beta \mu },\ ^{\circ }F_{\beta \mu }$
and $\ ^{\eta }\mathbf{F}_{\beta \mu }$ are computed by substituting  (\ref%
{ans1a}) and (\ref{ans2a}) into (\ref{gaugestr}). Any antisymmetric $\mathbf{%
F}_{\beta \mu }$ (\ref{gaugstr1}) is a solution of $D_{\mu }(\sqrt{|g|}%
F^{\mu \nu })=0,$ i.e.  determines $\ ^{\eta }F_{\beta \mu },\ ^{\eta }A_{\mu }$, for any given $\ ^{\circ }A_{\mu
},\ ^{\circ }F_{\beta \mu }.$ For nonholonomic modifications of scalar
fields, we take $\ \ ^{\circ }\phi (x^{1})\rightarrow \phi (x^{i},y^{a})=\
^{\phi }\eta (x^{i},y^{a})\ ^{\circ }\phi (x^{1})$ . It is supplemented with   a polarization $\
^{\phi }\eta $ for which
\begin{equation}
D_{\mu }\phi =0\mbox{ and \ }\phi (x^{i},y^{a})=\pm \phi _{\lbrack 0]}.
\label{cond3}
\end{equation}%
This nonholonomic configuration of the nonlinear scalar field is non-trivial
even with respect to N--adapted frames $\ ^{e}V(\phi )=0$ and $\
^{F}T_{\beta \delta }=0,$  (\ref{source1}). For ansatz (\ref{dm}), the
equations (\ref{cond3}) are%
\begin{equation}
(\partial /\partial x^{i}-A_{i})\phi =n_{i}\partial _{3}\phi +w_{i}\partial
_{t}\phi ,\ \left( \partial _{3}-A_{3}\right) \phi =0,\ \left( \partial
_{4}-A_{4}\right) \phi =0.  \notag
\end{equation}%
A nonholonomically deformed scalar (Higgs field depending in non--explicit
form on two variables because of constraint (\ref{cond3})) modifies
indirectly the off--diagonal components of the metric via $n_{i}$, $w_{i}$
and the above conditions for $\ ^{\eta }A_{\mu }.$

The effective gauge field $\mathbf{F}_{\beta \mu }$ (\ref{gaugstr1}) with
the potential $A_{\mu }$ (\ref{ans2a}) modified nonholonomically by $\phi $
and subject to the conditions (\ref{cond3}) determine exact solutions of the
system (\ref{eq2}) if the spacetime metric is chosen to be in the form (\ref%
{dm}). The energy--momentum tensor is determined to be  $\ ^FT_{\beta }^{\alpha
}=-4s^{2}\delta _{\beta }^{\alpha }$  \cite{lidsey}.
Interacting gauge and Higgs fields, with respect to N--adapted frames,
result in an effective cosmological constant $\ ^{s}\Lambda =8\pi s^{2}$
which should be added to the respective source (\ref{source}).

To conclude, a generic off--diagoanal ansatz $\widehat{\mathbf{g}}=[\eta _{i}\ ^{\circ
}g_{i},\eta _{a}\ ^{\circ }h_{a};w_{i},n_{i}]$ (\ref{dm}) and (effective)
gauge--scalar configurations $(A,\phi )$ subject to conditions mentioned
above define a decoupling of the nonlinear PDEs (\ref{ym1})--(\ref{heq3}) if
the sources (\ref{source}) are transformed in the form
\begin{equation}
\mathbf{\Upsilon }_{~\delta }^{\beta }=diag[\mathbf{\Upsilon }_{\alpha
}]\rightarrow \mathbf{\Upsilon }_{~\delta }^{\beta }+\ ^{F}T_{~\delta
}^{\beta }=diag[\mathbf{\Upsilon }_{\alpha }-4s^{2}\delta _{\beta }^{\alpha
}].  \label{ymsourc}
\end{equation}%
This is in sharp contrast to the situation where with respect to coordinate frames, such systems of equations describe a very complex, nonlinearly coupled gravitational and gauge--scalar interactions.

\subsection{Effective vacuum EYMH configurations in TMTs}

\label{ssvacuum}The effects of off--diagonal gravitational, scalar and gauge matter
fields result in  driving the vacuum energy density to zero even when
the effective source $\mathbf{\Upsilon }_{\alpha }$ and cosmological
constant $\Lambda _{0}$ are nontrivial. This is possible due to the
contributions of effective self--dual gauge fields. Such an effect is
discussed in \cite{mli} for instantons. If $\mathbf{\Upsilon }_{~\delta
}^{\beta }=0$ in (\ref{ymsourc}),   one  imposes further nonholonomic rescaling $%
\mathbf{\Upsilon \rightarrow }\Lambda _{0}$ when $ \Lambda _{0}-4s^{2}=0.$  We
can generate a very large class of solutions in TMTs with effective EMYH
interactions into nonholonomic vacuum configurations of modified
Einstein gravity. In this section, we analyze a subclass of generic
off--diagonal EYMH interactions which can be encoded as effective vacuum
Einstein manifolds of  various class and lead to solutions with nontrivial cosmological
constant $\Lambda _0 {=4s^{2}}.$ In general, such  solutions depend
parametrically on $\Lambda _{0}-4s^{2}$ and do not have a smooth limit from
non-vacuum to vacuum models. Effects of this type exist both in commutative,
noncommutative gauge gravity theories \cite{vnceym}, Einstein gravity
and its various modifications \cite{vhep,vnb15}, and TMTs. Examples are   provided
in the following sections.

Einstein equations (\ref{e1})--(\ref{confeq})
corresponding to a system of nonlinear PDEs (\ref{ym1})--(\ref{heq3}) with
source $\mathbf{\Upsilon }_{~\delta }^{\beta }$ (\ref{ymsourc}) are,
\begin{eqnarray}
\psi ^{\bullet \bullet }+\psi ^{\prime \prime } &=&2(\Upsilon -4s^{2}),
\label{ep1} \\
\partial _{t}\varpi \ \partial _{t}h_{3} &=&2h_{3}h_{4}(\Upsilon -4s^{2}),
\label{ep2} \\
\partial _{tt}^{2}n_{i}+\gamma \partial _{t}n_{i} &=&0,  \label{ep3} \\
\beta w_{i}-\alpha _{i} &=&0,  \label{ep4} \\
\partial _{i}\omega -n_{i}\partial _{3}\omega -(\partial _{i}\varpi
/\partial _{t}\varpi )\partial _{t}\omega &=&0.  \label{epconf}
\end{eqnarray}%
To derive self--consistent solutions of this system for $\Upsilon -4s^{2}=0$
we consider  off--diagonal ansatz depending on all spacetime coordinates,
\begin{eqnarray}
\widehat{\mathbf{g}} &=&e^{\psi (x^{k})}[dx^{1}\otimes dx^{1}+dx^{2}\otimes
dx^{2}]+h_{3}(x^{k},t)\underline{h}_{3}(x^{k},y^{3})\mathbf{e}^{3}\otimes
\mathbf{e}^{3}+h_{4}(x^{k},t)\mathbf{e}^{4}\otimes \mathbf{e}^{4},  \notag \\
\ \mathbf{e}^{3} &=&dy^{3}+n_{i}(x^{k})dx^{i},\mathbf{e}%
^{4}=dt+w_{i}(x^{k},t)dx^{i},  \label{sol2}
\end{eqnarray}%
where the coefficients of this target metric are defined by solutions of the
the following equations,
\begin{eqnarray}
\ddot{\psi}+\psi ^{\prime \prime } &=&0,  \label{ep1a} \\
(\partial _{t}\varpi )\ \partial _{t}h_{3} &=&0,  \label{ep2a} \\
\beta w_{i}-\alpha _{i} &=&0,  \label{ep4a}
\end{eqnarray}%
The coefficients $\beta $ and $\alpha _{i}$ are computed following
formulae (\ref{abc}) for nonzero $\partial _{t}\varpi \ $\ and $\partial
_{t}h_{3}\ $. The coefficients $h_{a}$, $\underline{h}_{3}$  and $w_{i}$ are additionally subject to the
zero--torsion conditions (\ref{lc}), (\ref{cosmc}) as in the form (\ref{lccond}) where, for simplicity, we fix $n_{i}$ equal to a constant as a trivial solutions of (\ref{ep3}).

For equation (\ref{ep1a}), we can take $\psi =0,$ or consider a trivial 2-d
Laplace equation with spacelike coordinates $x^{k}.$ There are two
possibilities to satisfy the condition (\ref{ep2a}) and derive the
corresponding off--diagonal solutions. In the first case we
take $h_{3}=h_{3}(x^{k})$, when $\partial _{t}h_{3}=0$. This implies
 that the equation (\ref{ep2a}) has solutions with zero source for
arbitrary function $h_{4}(x^{k},t)$ and arbitrary N--coefficients $%
w_{i}(x^{k},t)$ as follows from (\ref{abc}). For such vacuum
LC--configurations, the functions $h_{4}$ and $w_{i}$ are
general and should be constrained only by the conditions (\ref%
{lccond}). This constrains substantially the class of admissible $w_{i}$ if $%
h_{3}$ depends only on $x^{k}$ (we can perform a similar analysis as in
subsection \ref{sslccond}). The corresponding quadratic line element is
\begin{eqnarray}
ds^{2} &=&g_{\alpha \beta }(x^{k},t)du^{\alpha }du^{\beta }=e^{\psi
(x^{k})}[(dx^{1})^{2}+(dx^{2})^{2}]+  \label{odsolcv1} \\
&&\omega ^{2}(x^{k},y^{3},t)[\underline{h}%
_{3}(x^{k},y^{3})h_{3}(x^{k})(dy^{3})^{2}+h_{4}(x^{k},t)\ \left( dt+\partial
_{i}\check{A}(x^{k},t)\ dx^{i}\right) ^{2}],  \notag
\end{eqnarray}%
where we introduce a function $\check{A}=\check{A}(x^{k},t)$ for which $%
w_{i}=\partial _{i}\check{A}$ satisfies $\partial
_{i}w_{j}=\partial _{j}w_{i}$ in (\ref{lccondb}) and $\omega $ is a solution of
\begin{equation*}
\partial _{i}\omega -(\partial _{i}\check{A})\ \partial _{t}\omega =0.
\end{equation*}

In the second case a very different class of (off-) diagonal solutions result if we
choose, after corresponding coordinate transformations, $\varpi =\ln \left\vert
\partial _{t}h_{3}/\sqrt{|h_{3}h_{4}|}\right\vert =\ ^{0}\varpi =const$ and $%
\partial _{t}\varpi =0.$ For such configurations, we can consider $\partial
_{t}h_{3}\neq 0$ and solve (\ref{ep2a}) for
\begin{equation}
\sqrt{|h_{4}|}=\ ^{0}h\ \partial _{t}(\sqrt{|h_{3}|}),  \label{rel1}
\end{equation}%
with $\ ^{0}h$ equals a non vanishing constant. Such v--metrics are generated by any $%
f(x^{i},t)$ satisfying $\partial _{t}f \neq 0,$ when
\begin{equation}
h_{3}=f^{2}\left( x^{i},t\right) \mbox{ and }h_{4}=-(\ ^{0}h)^{2}\ \left[
\partial _{t}f\left( x^{i},t\right) \right] ^{2},  \label{aux2}
\end{equation}%
where the signs are  fixed in such a way that for $N_{i}^{a}\rightarrow 0$ we
obtain diagonal metrics with signature $(+,+,+,-).$ The coefficients (\ref%
{abc}) for (\ref{ep4a}) became trivial if $\alpha _{i}=\beta =0,$ and $%
w_{i}(x^{k},t)$ is any functions solving (\ref{lccond}). The last system
of equations for the LC--conditions are equivalent to
\begin{eqnarray}
\ \partial _{t}w_{i} &=&2\mathbf{\partial }_{i}\ln |f|-2w_{i}\partial
_{t}(\ln |f|),  \label{cond1} \\
\partial _{k}w_{i}-\mathbf{\partial }_{i}w_{k} &=&2(w_{k}\partial
_{i}-w_{i}\partial _{k})\ln |f|,  \notag
\end{eqnarray}%
for any $n_{i}(x^{k})$ when $\partial _{i}n_{k}=\partial _{k}n_{i}.$
Constraints of type $n_{k}\partial _{3}\underline{h}_{3}=\mathbf{\partial }%
_{k}\underline{h}_{3}$ have to be imposed for a nontrivial multiple $%
\underline{h}_{3}$ depending on $y^{3}.$

The corresponding quadratic line element is
\begin{eqnarray}
ds^{2} &=&g_{\alpha \beta }(x^{k},t)du^{\alpha }du^{\beta }=e^{\psi
(x^{k})}[(dx^{1})^{2}+(dx^{2})^{2}]+  \label{odsolcv2} \\
&&\omega ^{2}(x^{k},y^{3},t)\left[ \underline{h}_{3}(x^{k},y^{3})f^{2}\left(
x^{i},t\right) (dy^{3})^{2}-(\ ^{0}h)^{2}\ \left[ \partial _{t}f\left(
x^{i},t\right) \right] ^{2}\ \left( dt+w_{i}(x^{k},t)\ dx^{i}\right) ^{2}%
\right] ,  \notag
\end{eqnarray}%
where $w_{i}$ are taken to solve the conditions (\ref{cond1})with $\partial
_{i}w_{j}=\partial _{j}w_{i}$ and $\omega $ is a solution of%
\begin{equation*}
\partial _{i}\omega -w_{i}\ \partial _{t}\omega =0.
\end{equation*}

We conclude that off--diagonal interactions in effective EYMH systems
result in vanishing  cosmological constant as is demonstrated in the
 general solutions  (\ref{odsolcv1}) and (\ref{odsolcv2}) presented  above  for
LC--configurations. Such constructions can be generalized to include inhomogeneous
effective vacuum configurations with nontrivial nonholonomically induced
torsion (\ref{dtors}). Effects of this nature exist in TMTs when the analogous EYMH
systems are described by an action with two measures (\ref{acttmt}) related
to an action (\ref{effectconf}) via a N--adapted conformal transform (\ref%
{rescalling}). Additionally, a subclass of cosmological solutions satisfying the
conditions (\ref{lc}) and (\ref{cosmc}) can be generated if, for instance,
we restrict the generating functions in (\ref{odsolcv2}) to satisfy  via frame/coordinate transforms $f^{2}\left( x^{i},t\right) \rightarrow f^{2}\left( t\right) ,w_{i}(x^{k},t)\ \rightarrow
w_{i}(t),\omega \rightarrow 1$ and the integration functions are changed
into integration constants.

\subsection{Examples of (off--)diagonal nonholonomic deformations of
cosmological metrics}

In this section, we present details on  how  AFDM are employed to construct a new class of
inhomogeneous and anisotropic cosmological solutions with target d--metrics $\widehat{\mathbf{g}}$ (\ref{dm}) , with certain well defined limits for $\eta _{\alpha }\rightarrow 1$,  to a primed metric $\ ^{\circ }\mathbf{g}$. These can be interpreted as conformal,  frame or coordinate transformations of  the well known metrics like FLRW,
Bianchi, Kasner, or another metric corresponding to a particular cosmological solution \cite{calog,sungcoles1,mukh,weinb}.

\subsubsection{Off--diagonal deformations of FLRW configurations in TMTs}
\label{ssoddflrw}
We show how N--anholonomic FLRW deformations can be constructed to define
three classes of generic off--diagonal cosmological solutions for modified
EYMH systems in TMTs. Similar models for one measure theories are presented in  in \cite%
{vcosms}.

\noindent \paragraph{FLRW metrics:} For convenience, we  introduce the necessary notations to describe the primed standard FLRW metric, when written in the diagonal form, is %
\begin{equation}
\ \ ^{F}\mathbf{\mathring{g}}\mathbf{=}a^{2}(t)\left( \frac{dr{\otimes }dr}{%
1-Kr^{2}}+r^{2}d\theta {\otimes }d\theta \right) +a^{2}(t)r^{2}\sin
^{2}\theta d\varphi {\otimes }d\varphi -dt{\otimes }dt,  \label{frw}
\end{equation}%
with $K=\pm 1,0$ and spherical coordinates $x^{1}=r,x^{2}=\theta ,y^{3}=\varphi ,y^{4}=t$ and 
\begin{equation*}
\ ^{F}\mathring{g}_{1}=a^{2}/(1-\kappa r^{2}),\ ^{F}\mathring{g}%
_{2}=a^{2}r^{2}/(1-\kappa r^{2}),\ \ ^{F}\mathring{h}_{3}=a^{2}(t)r^{2}\sin
^{2}\theta ,\ \ ^{F}\mathring{h}_{4}=-1,\ \ ^{F}\mathring{N}_{i}^{a}=0.
\end{equation*}%
For simplicity, we take $K=0$ and choose Cartesian coordinates $%
(x^{1}=x,x^{2}=z,y^{3}=y,y^{4}=t),$ when the coefficients of  $ \ ^{F}%
\mathbf{\mathring{g}}$ are taken, respectively, in the form $\ ^{F}\mathring{%
g}_{1}=$ $\ ^{F}\mathring{g}_{2}=a^{2},\ \ ^{F}\mathring{h}_{3}=a^{2}\ \
_{F}h_{4}=-1$ and $\ ^{F}\mathring{N}_{i}^{a}=0.$ In this case, the
nontrivial coefficients of the primed diagonal metric depend only on the time
like coordinate $t$ and takes the form,
\begin{equation}
\ \ \ ^{F}\mathbf{\mathring{g}=}a^{2}(t)\left( dx{\otimes }dx+dz{\otimes }%
dz+dy{\otimes }dy\right) -dt{\otimes }dt.  \label{frw1}
\end{equation}%
Here we also note that instead of FLRW we can consider any other 'primed' metric $\
^{\circ }\mathbf{g}$ that can be a Bianchi, Kasner or a metric of a particular cosmological solution \cite%
{sungcoles1,pontzen,odintsov,capozz,linde4}.

The metrics (\ref{frw}) and/or (\ref{frw1}) define exact homogeneous
cosmological solutions of equations (\ref{enstdt}) and (\ref{lccond}) with
source $\mathbf{\Upsilon }_{\alpha \beta }=\frac{\kappa }{2}\ T_{\alpha
\beta }$ for a perfect fluid energy--momentum stress tensor,%
\begin{equation}
T_{\ \ \beta }^{\alpha }=diag[-p,-p,\rho ,-p].  \label{fluid}
\end{equation}%
Here $\rho $ and $p$ are the proper energy density and pressure
in the fluid rest frame. The Einstein equations corresponding to ansatz (\ref%
{frw}) take the form of two coupled nonlinear ODE (the Friedmann equations)%
\begin{equation}
H^{2}\equiv \left( \frac{\partial _{t}a}{a}\right) ^{2}=\frac{1}{3}\rho -%
\frac{\kappa }{a^{2}}  \label{fr1}
\end{equation}%
and
\begin{equation}
\partial _{t}H+H^{2}=\frac{\partial _{tt}^{2}a}{a}=-\frac{1}{6}(\rho +3p).
\label{fr2}
\end{equation}
The Hubble constant $H\equiv \partial _{t}a/a$ has the
units of inverse time and is positive (negative) for an expanding
(collapsing) universe. The equations (\ref{fr1}) and (\ref{fr2}) are related
via the condition $\nabla _{\alpha }T_{\ \ \beta }^{\alpha }=0,$  for which the
considered diagonal homogeneous ansatz is written as
\begin{equation*}
\partial _{t}\rho +3H(\rho +p)=0.
\end{equation*}%
Here we note that the strong energy conditions for matter, $\rho +3p\geq 0,$ or equivalently, the equation of state, $w=p/\rho \geq -1/3$, must be satisfied for an expanding universe.

\noindent \paragraph{Off--diagonal effective EYMH cosmological solutions of type 1:}
\vskip .3in

In this case the d--metric is of the type (\ref{nkillingsol}) with $%
\partial _{t}h_{a}\neq 0,\partial _{t}\varpi \neq 0$ and $\Upsilon
-4s^{2}\neq 0,$ when

\begin{equation*}
h_{3}=\frac{\ ^{s}\widetilde{\Psi }^{2}}{4(\Lambda _{0}-4s^{2})}=\eta _{3}\
^{F}\mathring{h}_{3}\mbox{ and
}h_{4}=\frac{(\partial _{t}\ ^{s}\widetilde{\Psi })^{2}}{\ ^{s}\Xi }=\eta
_{4}\ ^{F}\mathring{h}_{4}
\end{equation*}%
correspond to an effective cosmological constant $\Lambda _{0}-4s^{2}\neq 0$
with re--defined generating functions, $\ ^{s}\Psi \longleftrightarrow \ ^{s}%
\widetilde{\Psi }.$ The left label "s" emphasizes that such values encode
contributions from effective gauge fields, where
\begin{equation}
\ ^{s}\Psi ^{2}=(\Lambda _{0}-4s^{2})^{-1}\int dt(\Upsilon -4s^{2})\partial
_{t}(\ ^{s}\widetilde{\Psi }^{2})\mbox{
and }\ ^{s}\widetilde{\Psi }^{2}=(\Lambda _{0}-4s^{2})\int dt(\Upsilon
-4s^{2})^{-1}\partial _{t}(\ ^{s}\Psi ^{2}).  \label{rescgfflrw}
\end{equation}%
The functional
\begin{equation*}
\ ^{s}\Xi \lbrack \Upsilon ,\ ^{s}\widetilde{\Psi }]=\int dt(\Upsilon
-4s^{2})\partial _{t}(\ ^{s}\widetilde{\Psi }^{2})
\end{equation*}%
in the formula for $h_{4}$ in (\ref{solha}) can be considered as a
re--defined source,$\ \Upsilon -4s^{2}\rightarrow \ ^{s}\Xi ,$ for a
prescribed generating function $\widetilde{\Psi },$ when $\Upsilon
-4s^{2}=\partial _{t}(\ ^{s}\Xi )/\partial _{t}(\ ^{s}\widetilde{\Psi }%
^{2}). $ Such effective sources contain information on effective EYMH
interactions in TMTs. For convenience we  work  with a couple of generating
data, $(\ ^{s}\Psi ,\ ^{v}\Lambda -4s^{2})$ and $[\ ^{s}\widetilde{\Psi }%
,\Lambda _{0}-4s^{2},\ ^{s}\Xi ]$ related by formulae (\ref{rescgfflrw}) for
a prescribed effective cosmological constant $\Lambda _{0}$ and the
parameter $s$ for gauge fields. Such values have to be fixed in a form which
is compatible with experimental/ observational data, and result in a small
vacuum density. Summarizing the results for off--diagonal nonholonomic
deformations of the prime metric (\ref{frw1}), we get a quadratic line
element
\begin{eqnarray}
ds^{2} &=&g_{\alpha \beta }(x^{k},t)du^{\alpha }du^{\beta }=e^{\psi
(x,z)}[(dx)^{2}+(dz)^{2}]  \label{nkillingoffflrw} \\
&&+(\ ^{s}\omega )^{2}\frac{\ ^{s}\widetilde{\Psi }^{2}}{4(\Lambda
_{0}-4s^{2})}[dy+\left( \ _{1}n_{k}(x,z)+_{2}\widetilde{n}_{k}(x,z)\int dt%
\frac{(\partial _{4}\ ^{s}\widetilde{\Psi })^{2}}{(\ ^{s}\widetilde{\Psi }%
)^{3}\ ^{s}\Xi }\right) dx^{k}]^{2}  \notag \\
&&+(\ ^{s}\omega )^{2}\frac{(\partial _{t}\ ^{s}\widetilde{\Psi })^{2}}{\
^{s}\Xi }\ [dt+\frac{\partial _{i}\ ^{s}\Xi }{\partial _{t}\ ^{s}\Xi }%
dx^{i}]^{2},  \notag
\end{eqnarray}%
where $\ ^{s}\omega (x,z,y,t)$ is a solution of (\ref{confeq}) which for our
data is written in the form
\begin{equation*}
\partial _{i}\omega -n_{i}\partial _{3}\omega -w_{i}\partial _{t}\omega =0.
\end{equation*}%
For the N--connection coefficients, we have
\begin{equation*}
n_{i}=\ _{1}n_{k}(x,z)+_{2}\widetilde{n}_{k}(x,z)\int dt\frac{(\partial
_{4}\ ^{s}\widetilde{\Psi })^{2}}{(\ ^{s}\widetilde{\Psi })^{3}\ ^{s}\Xi }%
\mbox{ and }w_{i}=\frac{\partial _{i}\ ^{s}\Xi }{\partial _{t}\ ^{s}\Xi }.
\end{equation*}%
The function $\psi (x,z)$ in (\ref{nkillingoffflrw}) is a solution of (\ref%
{ep1}), i.e. of $\partial _{xx}^{2}\psi +\partial _{zz}^{2}\psi =2(\Upsilon
-4s^{2}).$

To understand possible physical implications of d--metrics (\ref%
{nkillingoffflrw}) it is more convenient to use the so--called polarization
functions $\eta _{\alpha }$ and $\eta _{i}^{a}:=N_{i}^{a}-\mathring{N}%
_{i}^{a}$ as in (\ref{dm}) and parameterize such solutions in the form
\begin{eqnarray}
\ \widehat{\mathbf{g}} &\mathbf{=}&\eta _{1}\ ^{F}\mathring{g}_{1}dx{\otimes
}dx+\eta _{2}\ ^{F}\mathring{g}_{2}dz{\otimes }dz+(\ ^{s}\omega )^{2}\left[
\eta _{3}\ ^{F}\mathring{h}_{3}\mathbf{e}^{3}{\otimes }\mathbf{e}^{3}+\eta
_{4}\ ^{F}\mathring{h}_{4}\mathbf{e}^{4}{\otimes }\mathbf{e}^{4}\right] ,
\notag \\
\mathbf{e}^{3} &=&dy+\eta _{1}^{3}dx+\eta _{2}^{3}dz,\mathbf{e}^{4}=dt+\eta
_{1}^{4}dx+\eta _{2}^{4}dz,  \label{nkillingoffflrw1}
\end{eqnarray}%
where
\begin{equation*}
\eta _{1}=\eta _{2}=a^{-2}(t)e^{\psi (x,z)},\eta _{3}=\ ^{s}\widetilde{\Psi }%
^{2}/4(\Lambda _{0}-4s^{2})a^{2}(t),\eta _{4}=(\partial _{t}\ ^{s}\widetilde{%
\Psi })^{2}/\ ^{s}\Xi ,\eta _{i}^{3}=n_{i},\eta _{i}^{4}=w_{i}
\end{equation*}%
are determined by the above solutions for the coefficients of the target
d--metric.

Solutions (\ref{nkillingoffflrw1}) describe general off--diagonal
deformations of the FLRW metrics in TMTs encoding modified EYMH
interactions. Such interactions may result in changing the topology and
symmetries, and are characterized by inhomogeneous, locally anistropic
configurations or non--perturbative effect.  The problem of physical
interpretation of such cosmological off--diagonal solutions is simplified to some extent
if we consider small deformations with polarizations of the type $\eta
_{\alpha }$ $\approx 1+\chi _{\alpha }$ and $\eta _{i}^{a} \approx 0+\chi
_{i}^{a}$ for small values $|\chi _{\alpha }|\ll 1$ and $|\chi _{i}^{a}|\ll
1,$ by which we obtain small deformations of the FLRW universes by certain
generalized two measure interactions and/or modified gravity theories with
effective EYMH fields. Nevertheless, even in such cases the target
configuration may encode nonlinear and nonholonomic parametric effects as
results of re--scaling (\ref{rescgfflrw}) of generating functions. This way
we model nonlinear nonholonomic transformations of a FLRW universe into an
effective and small--deformed one with small values of effective
cosmological constant, nonlinear  anisotropic processes and other effects of similar magnitude.

\noindent \paragraph{Off--diagonal cosmological solutions of type 2 and "losing"
information on effective EYMH:}
\vskip .3in

This class of solutions are characterized by the condition $\partial
_{t}h_{3}=0.$ The equation (\ref{ep2}) can be solved only if $\Upsilon
-4s^{2}=0,$ i.e. when the contributions from effective YM fields compensate
other (effective) modified gravity and/or matter field sources. We take
 the function $w_{i}(x^{k},t)$ as a  solution of (\ref{ep4}), or  its
equivalent (\ref{ep4a}), because the coefficients $\beta $ and $\alpha _{i}$
from (\ref{abc}) are zero. To find nontrivial values of $n_{i}$ we
integrate (\ref{ep3}) for $\partial _{t}h_{3}=0$ for any given $h_{3}$ and
find $n_{i}=\ ^{1}n_{k}\left( x^{i}\right) +\ ^{2}n_{k}\left( x^{i}\right)
\int h_{4}dt.$ Also, we take $g_{1}=g_{2}=e^{\psi (x^{k})},$ with $%
\psi (x^{k})$ determined by (\ref{ep1}) for a given source $(\Upsilon
-4s^{2}).$

In summary, this class of solutions can be chosen to be defined by the ansatz
\begin{eqnarray}
\widehat{\mathbf{g}} &\mathbf{=}&e^{\psi (x^{k})}{dx^{i}\otimes dx^{i}}+\
^{0}h_{3}(x^{k})\mathbf{e}^{3}{\otimes }\mathbf{e}^{3}+h_{4}(x^{k},t)\mathbf{%
e}^{4}{\otimes }\mathbf{e}^{4},  \label{genans1} \\
\mathbf{e}^{3} &=&dy+\left[ \ ^{1}n_{k}\left( x^{i}\right) +\
^{2}n_{k}\left( x^{i}\right) \int h_{4}dt\right] dx^{i},\mathbf{e}%
^{4}=dt+w_{i}(x^{k},t)dx^{i},\   \notag
\end{eqnarray}%
for arbitrary generating functions $h_{4}(x^{k},t),w_{i}(x^{k},t),\
^{0}h_{3}(x^{k})$ and integration functions $\ ^{1}n_{k}\left( x^{i}\right) $
and $\ ^{2}n_{k}\left( x^{i}\right)$. In general, such solutions carry
nontrivial nonholonomically induced torsion (\ref{dtors}).

The conditions (\ref{lccond}) constrains (\ref{genans1}) to a subclass of
LC--solutions resulting in the following equations
\begin{eqnarray}
\ ^{2}n_{k}\left( x^{i}\right) =0\ &\mbox{ and }&\partial _{i}\
^{1}n_{k}=\partial _{k}\ ^{1}n_{i},  \notag \\
\partial _{t}w_{i}+\partial _{i}(\ ^{0}h_{3})=0 &\mbox{ and }&\partial _{i}\
w_{k}=\partial _{k}\ w_{i},  \label{lc01}
\end{eqnarray}%
for any  $w_{i}(x^{k},t)$ and $\ ^{0}h_{3}(x^{k}).$ This class of
constraints on solutions (\ref{genans1}) do not involve the generating
function $h_{4}(x^{k},t)$ but only the N--connection coefficients for a
prescribed value $\ ^{0}h_{3}(x^{k})$.

Another metric to consider is the prime FLRW metric as in (\ref{frw}) and/or (\ref{frw1}) and repeat the constructions for the metric (\ref{nkillingoffflrw1}) but with the difference that we take  $\partial _{t}h_{3}=0.$ However, we study here another possibility, i.e., to begin with a prime metric which is not a solution of gravitational field equations and finally to generate off--diagonal cosmological solution with effective nontrivial nonholonomic vacuum configuration. Let us consider $\ ^{\circ}g_{i}=1,^{\circ }h_{3}=1,\ ^{\circ }h_{4}(t)=-a^{-2}(t)$,\ which, by TMT with effective EYMH anisotropic and inhomogeneous nonlinear interactions, result in target d--metrics of the type (\ref{genans1}). Using polarization functions we write
\begin{equation*}
\eta _{i}=e^{\psi (x^{k})},\eta _{3}=\ \ ^{0}h_{3}(x^{k}),\eta
_{4}=h_{4}(x^{k},t)a^{2}(t),\eta _{i}^{3}=\ ^{1}n_{k}\left( x^{i}\right) +\
^{2}n_{k}\left( x^{i}\right) \int h_{4}dt,\eta _{i}^{4}=w_{i}(x^{k},t),
\end{equation*}%
with $\ ^{\circ }w_{i}(t)=0$ and $\ ^{\circ }n_{i}(t)=0.$ Such cosmological solutions are constructed  as nonholonomic deformations of a conformal transformation  (with multiplication on factor $a^{-2}(t)$) of  the FLRW metric (\ref{frw}). We work with polarization functions $\eta _{4}(x^{k},t)$ when $h_{4}=\eta _{4}\ ^{\circ }h_{4}(t)\rightarrow $ $-a^{-2}(t)h_{4}(x^{k},t)$ for $\eta _{4}\rightarrow 1.$ The solutions are written in the form
\begin{eqnarray}
\ \widehat{\mathbf{g}} &\mathbf{=}&\eta _{1}\ dx{\otimes }dx+\eta _{2}\ dz{%
\otimes }dz+(\ ^{s}\omega )^{2}\left[ \eta _{3}\mathbf{e}^{3}{\otimes }%
\mathbf{e}^{3}-\eta _{4}\ a^{-2}(t)\mathbf{e}^{4}{\otimes }\mathbf{e}^{4}%
\right] ,  \notag \\
\mathbf{e}^{3} &=&dy+\eta _{1}^{3}dx+\eta _{2}^{3}dz,\mathbf{e}^{4}=dt+\eta
_{1}^{4}dx+\eta _{2}^{4}dz,  \label{nkillingoffflrw2}
\end{eqnarray}%
where
\begin{equation*}
\eta _{1}=\eta _{2}=a^{-2}(t)e^{\psi (x,z)},\eta _{3}=a^{2}(t)(\
^{0}h_{3}),\eta _{4}=h_{4}a^{2},\eta _{i}^{3}=n_{i},\eta _{i}^{4}=w_{i}
\end{equation*}%
are  the coefficients of the target d--metric (\ref{genans1}).

The class of solutions (\ref{nkillingoffflrw1}) represent the off--diagonal deformations of the FLRW metrics in TMTs encoding effective gauge and scalar field interactions when the effective cosmological constant is fixed to be
zero. We  generate  solutions with non-Killing symmetry for nontrivial  $v$--conformal factors $\ ^{s}\omega (x,z,y,t)$ subject to the constraints
\begin{equation*}
\partial _{i}\ ^{s}\omega -\eta _{i}^{3}\ (\partial _{3}\ ^{s}\omega) -\eta
_{i}^{4}\ (\partial _{t}\ ^{s}\omega) =0.
\end{equation*}%
The LC--conditions (\ref{lc01}) constraints substantially the time dependence of $\eta _{i}^{4}=w_{i}(x^{k},t).$ The class of solutions with nontrivial nonholonomic torsion (\ref{dtors}) allow arbitrary dependencies
on $t$ for N--connection coefficients $w_{i}.$

Off--diagonal metrics (\ref{genans1}) \ result only with time like dependence in the coefficients i.e.,  when $h_{4}=h_{4}(t),  \,\,\,w_{i}=w_{i}(t)$ and $n_{i}(t)$ are determined with some constant values of $\ ^{0}h_{4},\ \
^{1}n_{k},\ ^{2}n_{k}.$ Such conditions are relevant   for the Levi--Civita configurations if $w_{i}=const.$ This defines solutions of the Einstein equations with nonholonomic vacuum encoding TMTs contributions and effective
EYMH interactions. They transform nonholonomically a FLRW universe into
certain effective vacuum Einstein configurations which in this particular
case is diagonalizable  by coordinate transformations.
\noindent \paragraph{Off--diagonal cosmological solutions of type 3 and effective matter fields interactions:}
\vskip .3in
Non--vacuum metrics with $\partial _{t}h_{3}\neq 0$ and $\partial _{t}h_{4}=0$ are generated by taking the ansatz
\begin{eqnarray}
\widehat{\mathbf{g}} &\mathbf{=}&e^{\psi (x^{k})}{dx^{i}\otimes dx^{i}+}%
h_{3}(x^{k},t)\ \mathbf{e}^{3}{\otimes }\mathbf{e}^{3}-\ ^{0}h_{4}(x^{k})\
\mathbf{e}^{4}{\otimes }\mathbf{e}^{4},  \notag \\
\mathbf{e}^{3} &=&dy^{3}+n_{i}(x^{k},t)dx^{i},\ \mathbf{e}%
^{4}=dt+w_{i}(x^{k},t)dx^{i},  \label{genans2}
\end{eqnarray}%
where $g_{1}=g_{2}=e^{\psi (x^{k})},$ where $\psi (x^{k})$ is a solution of (\ref{eq1b}) for any given $\ ^{v}\Upsilon (x^{k})-4s^{2}.$ The function $h_{3}(x^{k},t)$ is constrained to satisfy the equation (\ref{eq2b}) which for $\partial _{t}h_{4}=0$ leads to
\begin{equation}
\partial _{tt}^{2}h_{3}-\frac{\left( \partial _{t}h_{3}\right) ^{2}}{2h_{3}}%
-2\ ^{0}h_{4}\ h_{3}[\Upsilon (x^{k},t)-4s^{2}]=0,  \label{aux01}
\end{equation}%
where the constant $s^{2}$ is introduced as an additional source in order to take into account possible contributions resulting from (anti) self--dual fields. The N--connection coefficients are
\begin{equation*}
w_{i}=\partial _{i}\widetilde{\Psi }/\partial _{t}\widetilde{\Psi },~n_{i}=\
^{1}n_{k}\left( x^{i}\right) +\ ^{2}n_{k}\left( x^{i}\right) \int [1/(\sqrt{%
|h_{3}|})^{3}]dt,
\end{equation*}%
where $\widetilde{\Psi }=\ln |\partial _{t}h_{3}/\sqrt{|\ ^{0}h_{4}h_{3}|}|.$

The Levi--Civita configurations for solutions (\ref{genans2}) are selected by the conditions (\ref{lccondb}) which, for this case, are satisfied if
\begin{equation*}
\ ^{2}n_{k}\left( x^{i}\right) =0\ \mbox{ and }\partial _{i}\
^{1}n_{k}=\partial _{k}\ ^{1}n_{i},
\end{equation*}%
and
\begin{eqnarray*}
\partial _{t}\left( w_{i}[\widetilde{\Psi }]\right) +w_{i}[\widetilde{\Psi }%
]\partial _{t}\left( h_{3}[\widetilde{\Psi }]\right) +\partial _{i}h_{3}[%
\widetilde{\Psi }] &=&0, \\
\partial _{i}\ w_{k}[\widetilde{\Psi }] &=&\partial _{k}\ w_{i}[\widetilde{%
\Psi }].
\end{eqnarray*}%
Such conditions are similar to (\ref{lc01}) but for a different relation of v--coefficients of d--metrics to another type of generating function $\widetilde{\Psi }.$ They are always satisfied for cosmological solutions
with $\widetilde{\Psi }=\widetilde{\Psi }(t)$ or if $\widetilde{\Psi }=const$ (in the last case $w_{i}(x^{k},t)$ can be any functions as follows from (\ref{eq3b}) with zero $\beta $ and $\alpha _{i},$ see (\ref{abc})).

\begin{eqnarray}
\ \widehat{\mathbf{g}} &\mathbf{=}&\eta _{1}\ ^{F}\mathring{g}_{1}dx{\otimes
}dx+\eta _{2}\ ^{F}\mathring{g}_{2}dz{\otimes }dz+(\ ^{s}\omega )^{2}\left[
\eta _{3}\ ^{F}\mathring{h}_{3}\mathbf{e}^{3}{\otimes }\mathbf{e}^{3}+\eta
_{4}\ ^{F}\mathring{h}_{4}\mathbf{e}^{4}{\otimes }\mathbf{e}^{4}\right] ,
\notag \\
\mathbf{e}^{3} &=&dy+\eta _{1}^{3}dx+\eta _{2}^{3}dz,\mathbf{e}^{4}=dt+\eta
_{1}^{4}dx+\eta _{2}^{4}dz,  \label{genans2a}
\end{eqnarray}%
where
\begin{equation*}
\eta _{1}=\eta _{2}=a^{-2}(t)e^{\psi (x,z)},\eta _{3}=\ ^{s}\widetilde{\Psi }%
^{2}/4(\Lambda _{0}-4s^{2})a^{2}(t),\eta _{4}=(\partial _{t}\ ^{s}\widetilde{%
\Psi })^{2}/\ ^{s}\Xi ,\eta _{i}^{3}=n_{i},\eta _{i}^{4}=w_{i}.
\end{equation*}%
Any solution $h_{3}(x^{k},t)$ of the equation (\ref{aux01}) generates a
d--metric or (\ref{genans2a}) which depends on parameter $(\Lambda
_{0}-4s^{2})\neq 0.$ The singular case with $\Lambda _{0}=4s^{2}$\ can be
described by a d--metric (\ref{genans2}) when $h_{3}$ is a solution of $%
\partial _{tt}^{2}h_{3}-\left( \partial _{t}h_{3}\right) ^{2}/2h_{3}=0.$ For
such configurations, we lose information about $\Lambda _{0}$ and $s^{2}$
but certain encodings of matter field interactions are possible in the function $%
\psi (x^{k})$ if the right side source $\partial _{xx}^{2}\psi +\partial
_{zz}^{2}\psi =2(\Upsilon -4s^{2})$ is changed to the nontrivial case $2(\
^{v}\Upsilon -4s^{2})$ for an N--adapted and anisotropic source$\
^{v}\Upsilon (x^{k}).$

Finally, we emphasize that off--diagonal deformations of FLRW metrics in
TMTs with effective EYMH interactions sources of the type $\Upsilon -4s^{2}$ can
be used for driving to zero an effective cosmological constant or for modeling
parametric transforms to configurations with small effective vacuum energy.

\subsubsection{Nonhomogeneous EYMH effects in Bianchi cosmology in TMTs}

\label{ssecbcosm}Spatially homogeneous but anisotropic relativistic cosmological models were constructed following the Bianchi classification corresponding to symmetry properties of their spatial hypersurfaces
 \cite{grish,ellis,sungcoles1}. \ Such cosmological metrics are parameterized by orthonormal tetrad (vierbein) bases $e_{\alpha ^{\prime \prime }}=e_{\ \alpha ^{\prime \prime }}^{\underline{\alpha }}\partial /\partial u^{%
\underline{\alpha }},$ if
\begin{equation}
\ \ ^{B}\mathbf{g}_{\alpha ^{\prime \prime }\beta ^{\prime \prime }}=\ \
^{B}e_{\ \alpha ^{\prime \prime }}^{\underline{\alpha }}\ \ ^{B}e_{\ \beta
^{\prime \prime }}^{\underline{\beta }}\ \ ^{B}g_{\underline{\alpha }%
\underline{\beta }}=diag[1,1,1,-1]  \label{bianchim}
\end{equation}%
and
\begin{equation*}
\left[ \ \ ^{B}e_{\alpha ^{\prime \prime }},\ \ ^{B}e_{\beta ^{\prime \prime
}}\right] =\ ^{B}w_{\ \alpha ^{\prime \prime }\beta ^{\prime \prime
}}^{\gamma ^{\prime \prime }}\left( t\right) \ \ ^{B}e_{\gamma ^{\prime \prime }},
\end{equation*}%
are satisfied and the  'structure constants' depend on time like variables,
\begin{equation}
\ \ ^{B}w_{\ \alpha ^{\prime \prime }\beta ^{\prime \prime }}^{\gamma
^{\prime \prime }}\left( t\right) =\ \epsilon _{\ \alpha ^{\prime \prime
}\beta ^{\prime \prime }\tau ^{\prime \prime }}n^{\tau ^{\prime \prime
}\gamma ^{\prime \prime }}\left( t\right) +\delta _{\beta ^{\prime \prime
}}^{\gamma ^{\prime \prime }}b_{\alpha ^{\prime \prime }}\left( t\right)
-\delta _{\alpha ^{\prime \prime }}^{\gamma ^{\prime \prime }}b_{\beta
^{\prime \prime }}\left( t\right) .  \label{bstrc}
\end{equation}%
The values $\ ^{B}w_{\ \alpha ^{\prime \prime }\beta ^{\prime \prime}}^{\gamma ^{\prime \prime }}\left( t\right) $ are determined by some diagonal tensor, $n^{\tau ^{\prime \prime }\gamma ^{\prime \prime }},$ and
vector, $b_{\alpha ^{\prime \prime }},$ fields used for the mentioned classification. Depending on parametrization of such tensor and vector objects, one constructs the so--called Bianchi universes which are either open or
closed similar to the homogeneous and isotropic FLRW case. With nontrivial limits from observational cosmology, there exist the so --called Bianchi $I,V,VII_{0},VII_{h}$ and $IX$ universes and their corresponding cosmologies.

The AFDM\ allows us to generalize any Bianchi metric $\ \ ^{B}\mathbf{g}%
_{\alpha ^{\prime \prime }\beta ^{\prime \prime }}$ (\ref{bianchim}) into
locally anisotropic solutions. As the first step, we  transform a set
of coefficients $\ ^{B}g_{\underline{\alpha }\underline{\beta }}(t)$ into the
 prime metric using frame transformations, $\ ^{B}\mathbf{\mathring{%
g}}_{\alpha \beta }=\ \ ^{B}e_{\ \alpha ^{\prime \prime }}^{\underline{%
\alpha }}\ \ ^{B}e_{\ \beta ^{\prime \prime }}^{\underline{\beta }}\ \
^{B}g_{\underline{\alpha }\underline{\beta }}.$ One also needs to solve certain
quadratic algebraic equations for $\ ^{B}e_{\ \alpha ^{\prime \prime }}^{%
\underline{\alpha }}$ in order to define frame coefficients depending on
the coordinate $t,$ and $\ ^{B}\mathbf{\mathring{g}}%
_{\alpha \beta }$ is parameterized as a prime metric
\begin{eqnarray}
\ ^{B}\mathbf{\mathring{g}}_{\alpha \beta } &\mathbf{=}&\ ^{B}\mathring{g}%
_{i}{dx^{i}\otimes dx^{i}}+\ ^{B}\mathring{h}_{a}(t)\ ^{B}\mathbf{\mathring{e%
}}^{a}{\otimes }\ ^{B}\mathbf{\mathring{e}}^{a},\   \label{bianchima} \\
\mathbf{\mathring{e}}^{3} &=&dy^{3}+\ ^{B}\mathring{n}_{i}(t)dx^{i},\
\mathbf{\mathring{e}}^{4}=dt+\ ^{B}\mathring{w}_{i}(t)dx^{i}.  \notag
\end{eqnarray}%
We  generalize these anisotropic homogeneous cosmological metrics to  generic off--diagonal locally anisotropic and inhomogeneous configurations defining cosmological solutions in TMTs with effective EYMH interactions.

The target ansatz is considered to be of the type (\ref{dm}),
\begin{equation*}
\widehat{\mathbf{g}}=[\eta _{i}\ \ ^{B}\mathring{g}_{i},(\ ^{s}\omega )^{2}\
\eta _{a}\ \ ^{B}\mathring{h}_{a};\ ^{B}\mathring{n}_{i}+\eta _{1}^{3},\ ^{B}%
\mathring{w}_{i}+\eta _{i}^{4}],
\end{equation*}
with prime data determined by coefficients of (\ref{bianchima}). We construct metrics $\widehat{\mathbf{g}}$ defining generic off--diagonal solutions of the nonholonomic EYMH system in TMTs, (\ref{ep1})--(\ref{epconf}) with source (\ref{ymsourc}), following the same procedure as in section \ref{sslccond}. In terms of polarization functions, such  solutions take the following form
\begin{eqnarray}
\ \widehat{\mathbf{g}} &\mathbf{=}&\eta _{1}\ ^{B}\mathring{g}_{1}dx^{1}{%
\otimes }dx^{1}+\eta _{2}\ ^{B}\mathring{g}_{2}dx^{2}{\otimes }dx^{2}+(\
^{B}\omega )^{2}\left[ \eta _{3}\ ^{B}\mathring{h}_{3}\mathbf{e}^{3}{\otimes
}\mathbf{e}^{3}+\eta _{4}\ ^{B}\mathring{h}_{4}\mathbf{e}^{4}{\otimes }%
\mathbf{e}^{4}\right] ,  \notag \\
\mathbf{e}^{3} &=&dy^{3}+(\ ^{B}\mathring{n}_{1}+\eta _{1}^{3})dx^{1}+(\ ^{B}%
\mathring{n}_{2}+\eta _{2}^{3})dx^{2},  \label{biancoffd} \\
\mathbf{e}^{4} &=&dt+(\ ^{B}\mathring{w}_{1}+\eta _{1}^{4})dx^{1}+(\ ^{B}%
\mathring{w}_{2}+\eta _{2}^{4})dx^{2}.  \notag
\end{eqnarray}%
The off--diagonal deformations of Bianchi metrics determined by sources $\Upsilon -4s^{2}\neq 0,$ and $\Lambda _{0}-4s^{2}\neq 0,$ with $\partial _{t}h_{a}\neq 0,\partial _{t}\varpi \neq 0$ are computed as%
\begin{equation*}
\ ^{B}g_{1}=\eta _{1}\ ^{B}\mathring{g}_{1}=e^{\psi (x^{k})},\ \
^{B}g_{2}=\eta _{2}\ ^{B}\mathring{g}_{2}=e^{\psi (x^{k})},
\end{equation*}%
for $\psi (x^{k})$ being a solution of the Poisson equation $\partial _{11}^{2}\psi +\partial _{22}^{2}\psi =2(\Upsilon -4s^{2});$

\begin{equation*}
\ ^{B}h_{3}=\eta _{3}\ ^{B}\mathring{h}_{3}=\frac{\ ^{B}\widetilde{\Psi }^{2}%
}{4(\Lambda _{0}-4s^{2})}\mbox{ and
}\ ^{B}h_{4}=\eta _{4}\ ^{B}\mathring{h}_{4}=\frac{(\partial _{t}\ ^{B}%
\widetilde{\Psi })^{2}}{\ ^{B}\Xi },
\end{equation*}%
are computed for an effective cosmological constant $\Lambda _{0}-4s^{2}\neq
0$ with generating function
\begin{equation*}
\ ^{B}\Psi ^{2}=(\Lambda _{0}-4s^{2})^{-1}\int dt(\Upsilon -4s^{2})\partial
_{t}(\ ^{B}\widetilde{\Psi }^{2})\mbox{
or }\ ^{B}\widetilde{\Psi }^{2}=(\Lambda _{0}-4s^{2})\int dt(\Upsilon
-4s^{2})^{-1}\partial _{t}(\ ^{B}\Psi ^{2}).
\end{equation*}%
We put the left label "B" in our formulae in order to emphasize that certain values contain information on prime metrics. For simplicity, we omit "s" even when gauge and Higgs fields  contributions are there.

The functional
\begin{equation*}
\ ^{B}\Xi \lbrack \Upsilon ,\ ^{B}\widetilde{\Psi }]=\int dt(\Upsilon
-4s^{2})\partial _{t}(\ ^{B}\widetilde{\Psi }^{2})
\end{equation*}%
can be considered as a re--defined source,$\ \Upsilon -4s^{2}\rightarrow \
^{B}\Xi ,$ for a prescribed generating function $\ ^{B}\widetilde{\Psi }$
for locally anisotropic and inhomogeneous Bianchi configurations, when $%
\Upsilon -4s^{2}=\partial _{t}(\ ^{B}\Xi )/\partial _{t}(\ ^{B}\widetilde{%
\Psi }^{2}).$ This allows to compute the N--connection coefficients
\begin{eqnarray}
\ ^{B}n_{k}(x^{k},t) &=&\ ^{B}\mathring{n}_{k}+\eta _{k}^{3}=\
_{1}n_{k}(x^{i})+_{2}\widetilde{n}_{k}(x^{i})\int dt\frac{(\partial _{t}\
^{B}\widetilde{\Psi })^{2}}{(\ ^{B}\widetilde{\Psi })^{3}\ ^{B}\Xi }%
\mbox{
and }  \label{bncoef} \\
\ ^{B}w_{i}(x^{k},t) &=&\ ^{B}\mathring{w}_{i}+\eta _{i}^{4}=\frac{\partial
_{i}\ ^{B}\Xi }{\partial _{t}\ ^{B}\Xi },  \notag
\end{eqnarray}%
which is constrained additionally to define LC--configurations following
the procedure described in section \ref{sslccond}.

The $v$--conformal factor $\ ^{B}\omega (x^{k},y^{3},t)$ is a solution of (%
\ref{confeq}) with coefficients (\ref{bncoef}) when
\begin{equation*}
\partial _{i}\ ^{B}\omega -\ ^{B}n_{i}\ \partial _{3}\ ^{B}\omega -\
^{B}w_{i}\ \partial _{t}\ ^{B}\omega =0.
\end{equation*}%
Having constructed an inhomogeneous locally anisotropic cosmological metric $%
\ \widehat{\mathbf{g}}(x^{k},t)$ (\ref{biancoffd}), we  consider
additional assumptions on generating and integration functions when the
coefficients are homogeneous but with nonholonomically deformed Bianchi
symmetries. This is possible if we chose at the end "pure" time dependencies
$\ ^{B}\widetilde{\Psi }(t),\Upsilon (t),\ ^{B}h_{a}(t),$ $w_{i}(t)$ and
constant values$\ ^{B}g_{k}$ and $\ ^{B}n_{i}.$

\subsubsection{Kasner type metrics}

Another class of anisotropic cosmological metrics is determined by the
Kasner solution and various generalizations \cite{frolov,roberts,steinh}. Such
 4--d metrics are written in the form%
\begin{equation}
\ \ ^{K}\mathbf{g}\mathbf{=}t^{2p_{1}}dx{\otimes }dx+t^{2p_{3}}dz{\otimes }%
dz+t^{2p_{2}}dy{\otimes }dy-dt{\otimes }dt,  \label{kasner}
\end{equation}%
with$\ ^{K}g_{1}=t^{2p_{1}},\ \ ^{K}g_{2}=t^{2p_{3}},\ \
^{K}h_{3}=t^{2p_{2}},\ \ ^{K}h_{4}=-1$ and$\ ^{K}N_{i}^{a}=0.$ The constants
$p_{1},p_{2},p_{3}$ define solutions of the vacuum Einstein equations if
the following  conditions are satisfied%
\begin{equation}
2\ ^{3}P=\ ^{2}P-\ ^{1}P,  \label{kasner1}
\end{equation}%
for $\left( \ ^{1}P\right) ^{2}=\left( p_{1}\right) ^{2}+\left( p_{2}\right)
^{2}+\left( p_{3}\right) ^{2},\ ^{2}P=p_{1}+p_{2}+p_{3},\
^{3}P=p_{1}p_{2}+p_{2}p_{3}+p_{1}p_{3}.$ Following the anholonomic
deformation method, we generalize such solutions to generic off--diagonal cosmological configurations as in section \ref{ssvacuum} when $\Upsilon =4s^{2}.$

The data for a primary metric are taken as $\ \mathring{g}_{1}=1,\ \mathring{g}%
_{2}=t^{2(p_{3}-p_{1})},\ \mathring{h}_{3}=t^{2(p_{2}-p_{1})},$ $\mathring{h}%
_{4}=-t^{-2p_{1}}$ and $\ \mathring{N}_{i}^{a}=0$ with constants $%
p_{1},p_{2} $ and $p_{3}$ considered for (\ref{kasner}). For simplicity, let
us analyze solutions with $p_{3}=p_{1}$ and consider an example when a
Kasner universe is generalized to locally anisotropic configurations
characterized with gravitational polarizations%
\begin{equation*}
\ \eta _{i}=1,\eta _{3}=f\left( x^{i},t\right) ,\eta _{4}=\ ^{0}h^{2}\ \left[
\partial _{t}f\left( x^{i},t\right) \right] ^{2},\eta
_{i}^{3}=n_{i}(x^{k},t),\eta _{i}^{4}=w_{i}(x^{i},t).
\end{equation*}%
For $h_{a}=\eta _{a}\ ^{\circ }h_{a}$ and $N_{i}^{a}=\eta _{i}^{a}+\ ^{\circ
}N_{i}^{a},$ the target metric is of type (\ref{odsolcv2}) generated for $%
\Upsilon =4s^{2},$
\begin{eqnarray}
\widehat{\mathbf{g}} &\mathbf{=}&{dx^{1}\otimes dx^{1}+dx^{2}\otimes dx^{2}}%
+f^{2}\left( x^{i},t\right) t^{-2p_{1}}\mathbf{e}^{3}{\otimes }\mathbf{e}%
^{3}-\ ^{0}h^{2}\ \left[ \partial _{t}f\left( x^{i},t\right) \right]
^{2}t^{-2p_{1}}\mathbf{e}^{4}{\otimes }\mathbf{e}^{4},  \notag \\
\mathbf{e}^{3} &=&dy^{3}+n_{k}\left( x^{i},t\right) dx^{i},\mathbf{e}%
^{4}=dt+w_{i}(x^{k},t)dx^{i},  \label{kasneran}
\end{eqnarray}%
where $w_{i}=w_{i}(x^{i},t)$ are arbitrary functions and
\begin{equation*}
\ n_{k}=\ ^{1}n_{k}(x^{i})+\ ^{2}n_{k}(x^{i})\int dt\left[ \partial _{t}\ln
|f\left( x^{i},t\right) |\right] ^{2}.
\end{equation*}%
The coefficient $h_{4}$ is determined by $h_{3}$ following formula $\sqrt{|h_{4}|}=\ ^{0}h\ \partial _{t}\sqrt{|h_{3}|}$ which holds true for $\eta _{a}$ for arbitrary generating function $f\left( x^{i},t\right) $ if $%
p_{2}=p_{1}.$ Additional constraints on $f\left(x^{i},t\right) $ are needed if the last condition is not satisfied. In the limit of trivial polarizations, this d--metric results into a conformally transformed metric (with factor $t^{2p_{1}})$ of the Kasner solution (\ref{kasner}). In general, such primed metrics are not a solution of the Einstein equations for the
Levi--Civita connection but it is possible to chose gravitational polarizations that generate vacuum off--diagonal Einstein fields even when the conditions of type (\ref{kasner1}) are not satisfied.

To generate homogeneous but anisotropic solutions we  eliminate dependencies on space coordinates and consider arbitrary $w_{i}=w_{i}(t)$ and constant $\ ^{1}n_{k}$ and $\ ^{2}n_{k},$ when
\begin{equation*}
n_{k}=\ ^{1}n_{k}+\ ^{2}n_{k}\int dt\left[ \partial _{t}\ln |f\left(
t\right) |\right] ^{2}.
\end{equation*}%
For LC--configurations, we take $\ ^{2}n_{k}=0$ and impose constraints of type (\ref{cond1}) on $w_{i}(t)$.

In a similar manner, we  construct various nonholonomic deformations of the
Kasner universes of types 1-3 and/or and  generalize them to solutions of type
(\ref{kasneran}).

\section{Effective TMT Large Field Inflation with $^{c}\protect\alpha $%
--Attractors}

\label{s5}We consider a broad class of (off-) diagonal attractor solutions that arise naturally in (modified) gravity theories and TMTs and define what we imply by natural inflationary models.  In this work, we study cosmological attractors as they are considered for cosmological models in Refs. \cite{linde1,linde2,linde3}. The use of the word ‘attractor’ need to be clarified  as  a similar term is widely used in the theory of dynamical systems, for certain equilibrium configurations with critical points in the phase space, i. e., critical points which are stable. Our use of the word attractor solutions  is in the same spirit as the authors of references \cite{linde1,linde2,linde3}. What the authors of those works mean by cosmological attractors (see, for instance, Ref. \cite{linde1}) can be stated in their own words: “Several large classes of theories have been found, all of which have the same observational predictions in the leading order in 1/N. We called these theories “cosmological attractors.” In our approach, the use of the word "attractor" is similar but in a more general context for generic off-diagonal solutions. Certain configurations in our work are determined by solutions, in general, with nonholonomically induced torsion and can be restricted to LC-configurations. Under such assumptions, these configurations appear again in other models under consideration by us. We group all such models as having "cosmological attractor configurations" since the configurations are common to these class of models. It is implicit that  such solutions satisfy the conditions for “standard” cosmological attractor configurations (in A. Linde and co-author sense) only for certain sub-classes of nonholonomic constraints when the models are determined by imposing  constraints on the  corresponding generating and integration functions and integration constants. For general nonholonomic constraints, such configurations do not define  cosmological attractor configurations in the sense of the above mentioned original works \cite{linde1,linde2,linde3} but positively can be considered to possess similar properties for small off-diagonal deformations (perturbations) of the metrics. The important point is that such models have the same observational predictions in the leading order of $1/N$.  In this section, we shall define and study cosmological attractor configurations for modified gravity theories in terms of a parameter $^{c}\alpha $ that determines the curvature and cutoff. ”Henceforth in the following, in  order to make our manuscript more transparent, wherever we use the word "attractor" in the text , it will simply imply that certain  class of theories and respective off-diagonal cosmological solutions are generated which, under specific conditions on the parameter space  lead to the same observational predictions.

\subsection{Nonholonomic conformal transforms and cosmological attractors}


Attractor type configurations are possible to construct for a certain classes of
nonlinear scalar potentials in (\ref{act1}).  We use the term "configuration" because in that formula and in formula (\ref{ea1}) there are considered N--elongated derivatives. The equations are written with respect to nonholonomic bases and for generalized Ricci scalar curvature. As such additional assumptions are necessary in order to extract a "standard " cosmological attractor considered, for instance, in \cite{linde1}.  To begin with, we take the effective potential\ (\ref{effp}),
\begin{equation}
\ ^{e}V=\ ^{q}V=q^{2}(\tanh \phi ),  \label{effpe}
\end{equation}
for an arbitrary function $q$ and study the model  with the lagrangian
\begin{equation}
\ _{q}^{1}L=-\frac{1}{\kappa }\widehat{R}(\mathbf{g})+\frac{1}{2}\mathbf{g}%
^{\mu \nu }\mathbf{e}_{\mu }\phi \ \mathbf{e}_{\nu }\phi -q^{2}(\tanh \phi ).
\label{qth}
\end{equation}%
The equations (\ref{ea1}) impose the condition $\ _{q}^{1}L=M=const$ \footnote{In
this section, we use natural units $1/\kappa =1/2$}. Attractor models are
usually constructed in terms of two fields. In addition to $\phi (u^{\mu })$
we consider a second field $\chi (u^{\mu }).$  The fields $(\phi ,\chi )$
are subject to additional nonholonomic constraints involving the
generating function $\Psi =e^{\varpi }$ (\ref{genf}), some  possible
re--definitions (\ref{rescgf}) of effective matter field sources $\Upsilon $
and the effective cosmological constant $\Lambda _{0}.$

The theory (\ref{qth}) is related to a class of models
\begin{equation}
\ _{\chi }^{1}L=\frac{1}{2}\left[ \mathbf{g}^{\mu \nu }(\mathbf{e}_{\mu
}\phi \ \mathbf{e}_{\nu }\phi -\mathbf{e}_{\mu }\chi \ \mathbf{e}_{\nu }\chi
)+(\phi ^{2}-\chi ^{2})\widehat{R}(\mathbf{g})+q^{2}(\phi /\chi )(\phi
^{2}-\chi ^{2})^{2}\right]  \label{eqth}
\end{equation}%
by gauge condition
\begin{equation}
\phi ^{2}-\chi ^{2}=1 . \label{gauge12}
\end{equation}
The Lagrange density $\ _{\chi }^{1}L$ posses a $SO(1,1)$ symmetry
which is deformed by the term $q^{2}(\phi /\chi ).$ In  turn, the
Lagrange density $\ _{q}^{1}L$ may restore the $SO(1,1)$ symmetry at a
critical point because for large $\phi $ there exist  asymptotic limits, $%
\tanh \phi \rightarrow \pm 1$ and $q^{2}(\tanh \phi )\rightarrow const$. The
terms proportional to $q^{2}$ can be transformed into effective sources and
cosmological constant via eventual re-scaling of generating functions. Employing
 self--duality for  gauge field configurations with source $\Upsilon -4s^{2}, $
(\ref{ymsourc}), and using the gauge (\ref{gauge12}) with $\breve{\phi}%
=\sinh \phi $ and $\breve{\chi}=\cosh \chi ,$ we can approximate $\ _{\chi }^{1}L$ to %
\begin{equation*}
\ _{\breve{\phi}}^{1}L=\frac{1}{2}\left[ -\widehat{R}(\mathbf{g})+\mathbf{g}%
^{\mu \nu }\mathbf{e}_{\mu }\breve{\phi}\ \mathbf{e}_{\nu }\breve{\phi}%
+(\Lambda _{0}-4s^{2})\right] .
\end{equation*}%
Another important property
of the Lagrange density $\ _{\chi }^{1}L$ is that for a fixed value $q=q_{0}$
there is local conformal invariance under N--adapted transforms
\begin{equation}
\mathbf{\tilde{g}}_{\mu \nu }=e^{-2\sigma (u)}\mathbf{g}_{\mu \nu },\tilde{%
\chi}=e^{\sigma (u)}\chi ,\tilde{\phi}=e^{\sigma (u)}\phi .
\label{conftrfields}
\end{equation}%
Such a theory describes anti--gravity if $\phi ^{2}-\chi ^{2}>0,$ i.e. $\chi
$ represents the cutoff for possible values of the scalar field $\phi .$

By identifying $\sigma $ from (\ref{conftrfields}) with $\widehat{\sigma }$ in (%
\ref{rescalling}) when $e^{-2\widehat{\sigma }(u)}=2U/(V+M)=\Phi /\sqrt{|%
\mathbf{g}_{\alpha \beta }|}$ we present a model of a TMT theory of the type (\ref{acttmt})
derived for the action
\begin{eqnarray}
S &=&\int d^{4}u\left[ \ _{q}^{1}L\Phi +\ ^{2}L\sqrt{|\mathbf{g}_{\alpha
\beta }|}+N\phi \varepsilon ^{\mu \nu \alpha \beta }F_{\mu \nu
}^{a}F_{\alpha \beta }^{a}\right]  \label{actatract} \\
&\simeq &\int d^{4}u\left[ \ \ _{\chi }^{1}L\Phi +\ ^{2}L\sqrt{|\mathbf{g}%
_{\alpha \beta }|}+N\phi \varepsilon ^{\mu \nu \alpha \beta }F_{\mu \nu
}^{a}F_{\alpha \beta }^{a}\right]  \notag \\
&\simeq &\int d^{4}u\left[ \ \ _{\breve{\phi}}^{1}L\Phi +\ ^{2}L\sqrt{|%
\mathbf{g}_{\alpha \beta }|}+N\phi \varepsilon ^{\mu \nu \alpha \beta
}F_{\mu \nu }^{a}F_{\alpha \beta }^{a}\right] .  \notag
\end{eqnarray}%
The explicit construction depends on the type of generating functions, conformal transforms, effective sources, asymptotic limits and gauge conditions employed in our theory. This way we construct  different
toy TMT models with EYMHs which for data $(\phi ,\chi )$ possess attractor properties and the parameters defining such attractors encode off--diagonal gravitational and (effective) matter field interactions. It is a very
difficult technical task to construct cosmological solutions in such theories. Nevertheless, transforming any variant (\ref{actatract}) into an effective gravitational theory (\ref{effectconf}) with source $\ ^{e}\mathbf{%
T}_{\alpha \beta }\rightarrow \ ^{q}\mathbf{T}_{\alpha \beta }$ (\ref{sdcurv}) corresponding to $\ ^{q}V$ contributions, (\ref{effpe}), the effective EYMH equations (\ref{ym1})--(\ref{heq3}) can be integrated in very general
forms following the AFDM. Their solutions depend on integration functions and integration constants.

It is very surprising that the formulas (\ref{eqth})--(\ref{actatract}) and their physical consequences are similar to those for holonomic models considered in Refs. \cite{linde1,linde2,linde3}. Our solutions encoding  cosmological attractor configurations were derived for a class of modified theories with generalized off--diagonal metrics and nonlinear and distinguished linear connections and contributions from EYMHs for different TMT modes. It is not obvious, that such nonlinear systems may have a similar cosmological attractor  behavior like in the original works with diagonal solutions. Generic off-diagonal models can be elaborated following our geometric techniques with N--adapted nonholonomic variables and splitting of corresponding systems of nonlinear PDEs.  In such variables, it is possible to generate new classes of inhomogeneous and anisotropic solutions. Our goal was to  find such classes of nonholonomic constraints and subclasses of generating and integration functions, and constants, when  solutions with "hat" values and TMT--EYMH contributions really preserve the main physical properties of cosmological attractors. This emphasizes the  general importance of the results on cosmological attractors in the cited works due to A. Linde and co-authors. Our main conclusion is that for a corresponding class of nonholonomic constraints, a cosmological attractor configuration may "survive" for very general off--diagonal and matter source deformations, in various classes of TMT theories and effective Einstein like ones encoding  modified gravity theories.
\subsection{Effective interactions and cosmological attractors}

We can fix different gauge conditions but obtain the same results. For instance, we can work with $\chi (x)=1$ instead of (\ref{gauge12}), and the scalar field $\check{\phi}.$  With respect to
N--adapted Jordan frames, the total Lagrangian is
\begin{equation*}
\ _{J}^{1}L=-\frac{1}{2}\widehat{R}(\ ^{J}\mathbf{g})\ (1-\check{\phi}^{2})+%
\frac{1}{2}\mathbf{g}^{\mu \nu }\mathbf{e}_{\mu }\check{\phi}\ \mathbf{e}%
_{\nu }\check{\phi}+q^{2}(\check{\phi})\ (\check{\phi}^{2}-1)^{2}.
\end{equation*}%

We change  the d--metric into a conformally equivalent metric with equivalent
Einstein frame formulation in terms of $\ ^{E}\mathbf{g}_{\mu \nu },$ when $%
\ ^{E}\mathbf{g}_{\mu \nu }=(1-\check{\phi}^{2})\ ^{J}\mathbf{g}_{\mu \nu }.$
 The Lagrangian $\ _{J}^{1}L$ transforms into $\ _{E}^{1}L$ where %
\begin{equation}
\ _{E}^{1}L=-\frac{1}{2}\widehat{R}(\mathbf{g})\ +\frac{1}{(1-\check{\phi}%
^{2})^{2}}\mathbf{g}^{\mu \nu }\mathbf{e}_{\mu }\check{\phi}\ \mathbf{e}%
_{\nu }\check{\phi}+q^{2}(\check{\phi}).  \label{actenstfr}
\end{equation}%
Equivalently, $\ _{E}^{1}L$ transforms into $\ _{q}^{1}L$ (\ref{qth}) if the
scalar fields are re--defined as follows,
\begin{equation*}
\frac{d\phi }{d\check{\phi}}=(1-\check{\phi}^{2})^{-1},\mbox{ i.e. }\check{%
\phi}=\tanh \phi .
\end{equation*}

In the theory $\ _{E}^{1}L$ (\ref{actenstfr})  there is an ultra violet (UV)
cutoff $\Lambda =1,$ i.e. $\Lambda =M_{p}$ in terms of the Planck mass and if
$\phi $ become greater than $1$ we get a TMT theory with antigravity. Such
models were studied in the literature \cite{linde4,linde5} and other papers before the concept of cosmological attractors was introduced.  Our main goal is to study how "nonholonomically deformed" cosmological attractors can be modelled by nonholonomic constraints, generating functions and effective sources in such a way  that there are satisfied the criteria for "standard" cosmological attractors to emerge in the sense of Refs. \cite{linde1,linde2,linde3}.   We do not consider in this work similar constructions with nonholonomic variables but only emphasize that certain antigravity effects can be modelled by off--diagonal gravitational interactions and effective polarization of physical constants \cite{vhd2013}.   The previous formulae show that $\phi $ becomes infinitely large if $\check{\phi}%
\rightarrow 1$ but the effects of the cutoff can be ignored if $|\check{\phi}|\ll 1$, when $\check{\phi}\approx \phi .$ Excluding some very singular behaviour near the boundary of the moduli space, the asymptotic behaviour of
$\ ^{q}V(\phi )$ (\ref{effpe}) at large $\phi $ is universal. This universality exists for TMT models with effective EYMH interactions as follows from above equivalence (under well--defined conditions) of theories (\ref{actatract}) and (\ref{effectconf}).

The goal of this section is to study cosmological effects of (in general, locally anisotropic and inhomogeneous) attractors parameterized by a constant $\ ^{c}\alpha $ $\lesssim O(1).$ Attractor configurations can be
introduced in several inequivalent ways. We will generalize the constructions following \cite{linde1} and analyze possible off--diagonal solutions determined by sources $\ ^{q}V(\phi )$ and $\ ^{c}\alpha $--parameters.

We consider the Lagrangian
\begin{equation*}
\ _{E}^{\alpha }L=-\frac{1}{2}\widehat{R}(\mathbf{g})\ +\frac{\ ^{c}\alpha }{%
(1-\tilde{\phi}^{2})^{2}}\mathbf{g}^{\mu \nu }\mathbf{e}_{\mu }\tilde{\phi}\
\mathbf{e}_{\nu }\tilde{\phi}+q^{2}(\tilde{\phi}).
\end{equation*}%
which is given also in the Einstein frame as (\ref{actenstfr}) but contains a cutoff  $\ ^{c}\alpha $. We label the scalar field as $\tilde{\phi} $ (instead of $\check{\phi}$) in order to emphasize that we shall analyze a special class of solutions with $\ ^{c}\alpha $--dependence. We obtain a $\ ^{c}\alpha $--attractor configuration by re-scaling the scalar field,
\begin{equation*}
\frac{d\tilde{\phi}}{d\check{\phi}}=(1-\check{\phi}^{2})^{-1},\mbox{ i.e. }%
\check{\phi}=\tanh \tilde{\phi},\mbox{ and/ or redefining }\frac{\tilde{\phi}%
}{\sqrt{\ ^{c}\alpha }}=\tanh \frac{\phi }{\sqrt{\ ^{c}\alpha }},
\end{equation*}%
which leads to  effective theories of the type
\begin{eqnarray*}
\ _{E}^{\alpha }L &=&-\frac{1}{2}\widehat{R}(\mathbf{g})\ +\frac{\
^{c}\alpha }{(1-\tilde{\phi}^{2}/\ ^{c}\alpha )^{2}}\mathbf{g}^{\mu \nu }%
\mathbf{e}_{\mu }\check{\phi}\ \mathbf{e}_{\nu }\check{\phi}+q^{2}(\frac{%
\check{\phi}}{\sqrt{\ ^{c}\alpha }}) \\
&=&-\frac{1}{2}\widehat{R}(\mathbf{g})\ +\frac{1}{2}\mathbf{g}^{\mu \nu }%
\mathbf{e}_{\mu }\phi \ \mathbf{e}_{\nu }\phi +q^{2}(\tanh \frac{\phi }{%
\sqrt{\ ^{c}\alpha }}),
\end{eqnarray*}%
with a shifted cutoff position at $\Lambda =\sqrt{\ ^{c}\alpha }.$
\vskip .3 in
\subsection{Off--diagonal attractor type cosmological solutions}
\vskip .3 in
As alluded to in the previous subsection, the asymptotic behaviour of $\ ^{q}V(\phi )$ (\ref{effpe}) at large $\phi $ is universal. This universality  allows to construct various classes of generic off--diagonal cosmological metrics in modified  models of gravity with effective EYMH interactions using the conformal factor transformation (\ref{rescalling}). This is possible even when the generating functions and sources are very different for different classes of effective matter field interactions with nonlinear scalar potentials. The goal of this section is to prove how $q$--terms of the type $\ ^{q}V(\phi ,\ ^{c}\alpha )$ for attractors are encoded in various classes of solutions
studied in previous section. This holds for any
\begin{equation*}
e^{-2\ ^{c}\widehat{\sigma }(u)}=2U/\left[ \ ^{q}V(\phi ,\ ^{c}\alpha )+M%
\right] =\ ^{c}\Phi /\sqrt{|\ ^{c}\mathbf{g}_{\alpha \beta }|},
\end{equation*}%
where the left label "c" indicates that certain values refer to attractor
configurations with $\ ^{c}\alpha $--scale. The physical cosmological
d--metric $\ ^{c}\mathbf{g}_{\alpha \beta }$ is computed to be
\begin{equation}
\ ^{c}\mathbf{g}_{\mu \nu }=e^{2\ ^{c}\widehat{\sigma }(u)}\ ^{c}\widehat{%
\mathbf{g}}_{\mu \nu }=\frac{\ ^{q}V(\phi ,\ ^{c}\alpha )+M}{2U}\ ^{c}%
\widehat{\mathbf{g}}_{\mu \nu }.  \label{attractcf}
\end{equation}%
Having computed $\ ^{c}\mathbf{g}_{\mu \nu }$ for the data $\left[ \ ^{c}%
\widehat{\mathbf{g}}_{\mu \nu },\ ^{q}V,M,U\right] ,$ we construct a
corresponding TMT model when the second measure is taken to be
\begin{equation}
\ ^{c}\Phi =\frac{2U}{\ ^{q}V(\phi ,\ ^{c}\alpha )+M}\sqrt{|\ ^{c}\mathbf{g}%
_{\alpha \beta }|}.  \label{sm}
\end{equation}%
Formulae (\ref{attractcf}) and (\ref{sm}) can be applied to generate
solutions for the TMT system (\ref{ea1}) - (\ref{eq2}) if $\ ^{c}\widehat{%
\mathbf{g}}_{\mu \nu }$ is known as an attractor cosmological metric  (in general,
nonhomogeneous and locally anisotropic) for effective EYMH interactions.

\subsubsection{Off--diagonal effective EYMH cosmological attractor solutions of type 1}

Using  (\ref{nkillingoffflrw1}) and (\ref{attractcf}), we construct families of generic off--diagonal cosmological attractor configurations with metrics
\begin{eqnarray}
\ ^{c}\mathbf{g} &\mathbf{=}&e^{2\ ^{c}\widehat{\sigma }(u)}\mathbf{\{}\eta
_{1}\ ^{F}\mathring{g}_{1}dx{\otimes }dx+\eta _{2}\ ^{F}\mathring{g}_{2}dz{%
\otimes }dz+(\ ^{c}\omega )^{2}\left[ \eta _{3}\ ^{F}\mathring{h}_{3}\mathbf{%
e}^{3}{\otimes }\mathbf{e}^{3}+\eta _{4}\ ^{F}\mathring{h}_{4}\mathbf{e}^{4}{%
\otimes }\mathbf{e}^{4}\right] \},  \notag \\
\mathbf{e}^{3} &=&dy+\eta _{1}^{3}dx+\eta _{2}^{3}dz,\mathbf{e}^{4}=dt+\eta
_{1}^{4}dx+\eta _{2}^{4}dz,  \label{sola1}
\end{eqnarray}%
where the gravitational polarizations and N--connection coefficients are computed to be
\begin{equation*}
\ ^{c}\eta _{1}=\ ^{c}\eta _{2}=a^{-2}(t)e^{\ ^{c}\psi (x,z)},\eta _{3}=\
^{c}\widetilde{\Psi }^{2}/4(\Lambda _{0}^{c}-4s^{2})a^{2}(t),\ ^{c}\eta
_{4}=(\partial _{t}\ ^{c}\widetilde{\Psi })^{2}/\ ^{c}\Xi ,\ ^{c}\eta
_{i}^{3}=n_{i},\ ^{c}\eta _{i}^{4}=w_{i}.
\end{equation*}%
The parameter $\ ^{c}\alpha $ contributes to all data defining such nonholonomic deformations of FLRW primary metric because it is included in the effective source when $\Upsilon \rightarrow \ ^{c}\Upsilon $ with $\
^{c}\Upsilon -4s^{2}\neq 0.$ The corresponding effective cosmological constant is labeled $\Lambda _{0}^{c}$ and satisfies the condition $\Lambda _{0}^{c}-4s^{2}\neq 0$ (for the class of solutions of type 1). As a  result, the generating functions is redefined to simplify the formulas, $\ ^{c}\Psi \longleftrightarrow \ ^{c}\widetilde{\Psi },$ with
\begin{equation*}
\ ^{c}\Psi ^{2}=(\Lambda _{0}^{c}-4s^{2})^{-1}\int dt(\ ^{c}\Upsilon
-4s^{2})\partial _{t}(\ ^{c}\widetilde{\Psi }^{2})\mbox{
and }\ ^{c}\widetilde{\Psi }^{2}=(\Lambda _{0}^{c}-4s^{2})\int dt(\
^{c}\Upsilon -4s^{2})^{-1}\partial _{t}(\ ^{c}\Psi ^{2}).
\end{equation*}%
The information on $\ ^{q}V(\phi ,\ ^{c}\alpha )$ is also contained in the functional
\begin{equation*}
\ ^{c}\Xi \lbrack \ ^{c}\Upsilon ,\ ^{c}\widetilde{\Psi }]=\int dt(\
^{c}\Upsilon -4s^{2})\partial _{t}(\ ^{c}\widetilde{\Psi }^{2}).
\end{equation*}%
It is considered as a re--defined effective source,$\ \ ^{c}\Upsilon -4s^{2}\rightarrow \ ^{c}\Xi ,$ for a prescribed generating function $\ ^{c}\widetilde{\Psi },$ for which $\ ^{c}\Upsilon -4s^{2}=\partial _{t}(\
^{c}\Xi )/\partial _{t}(\ ^{c}\widetilde{\Psi }^{2}).$

We  express (\ref{sola1}) as a d--metric (\ref{dm}) with coefficients relevant to  the $v$--metric:

\begin{equation*}
h_{3}=\frac{\ ^{c}\widetilde{\Psi }^{2}}{4(\Lambda _{0}^{c}-4s^{2})}=\
^{c}\eta _{3}\ ^{F}\mathring{h}_{3}\mbox{ and
}h_{4}=\frac{(\partial _{t}\ ^{c}\widetilde{\Psi })^{2}}{\ ^{c}\Xi }=\
^{c}\eta _{4}\ ^{F}\mathring{h}_{4}.
\end{equation*}%
For the off--diagonal attractor N--connection coefficients, we compute
\begin{equation*}
n_{i}=\ _{1}n_{k}(x,z)+_{2}\widetilde{n}_{k}(x,z)\int dt\frac{(\partial
_{t}\ ^{c}\widetilde{\Psi })^{2}}{(\ ^{c}\widetilde{\Psi })^{3}\ ^{c}\Xi }%
\mbox{ and }w_{i}=\frac{\partial _{i}\ ^{c}\Xi }{\partial _{t}\ ^{c}\Xi }.
\end{equation*}

The "vertical" conformal factor $\ ^{c}\omega (x,z,y,t)$ is a solution of (\ref{confeq}) for which  attractor data is written in the form
\begin{equation*}
\partial _{i}\ ^{c}\omega -n_{i}\ \partial _{3}\ ^{c}\omega -w_{i}\ \partial
_{t}\ ^{c}\omega =0.
\end{equation*}

The function $\ ^{c}\psi (x,z)$ presented in the attractor's polarization functions is a solution of (\ref{ep1}) when $\partial _{xx}^{2}\ ^{c}\psi +\partial _{zz}^{2}\ ^{c}\psi =2(\ ^{c}\Upsilon -4s^{2}).$

Finally, we conclude that the formulae for the coefficients of the d--metric (\ref{sola1}) depend on the type of N--adapted frame and coordinate transforms necessary to fix  observational data. The conformal factor $%
e^{2\ ^{c}\widehat{\sigma }(u)}$ encodes attractor parameters in a more direct form.
\subsubsection{Generalized locally anisotropic Bianchi attractors}


Sources with attractor potential $\ ^{q}V(\phi )$ (\ref{effpe}) induce generic off--diagonal cosmological solutions, in general, with inhomogeneity and local anisotropy. For a target ansatz of type (\ref{dm}), we parameterize
\begin{equation*}
\ ^{c}\widehat{\mathbf{g}}=e^{2\ ^{c}\widehat{\sigma }}\ ^{c}\mathbf{g}=[\
^{c}\eta _{i}\ \ ^{B}\mathring{g}_{i},(\ ^{c}\omega )^{2}\ \ ^{c}\eta _{a}\
\ ^{B}\mathring{h}_{a};\ ^{B}\mathring{n}_{i}+\ ^{c}\eta _{1}^{3},\ ^{B}%
\mathring{w}_{i}+\ ^{c}\eta _{i}^{4}],
\end{equation*}%
when the prime solution $\ ^{B}\mathbf{\mathring{g}}$ is determined by coefficients of (\ref{bianchima}). Our purpose is to state the conditions when $\ ^{c}\mathbf{g}$ from above formula defines generic off--diagonal
solutions with attractor properties in TMTs with effective EYMH interactions, i.e. of (\ref{ep1})--(\ref{epconf}) with source (\ref{ymsourc}) encoding an attractor potential.\footnote{It is supposed that the parameter $\ ^{c}\alpha $  contributes to all data defining nonholonomic deformations of a primary Biachi metric.  This parameter is included into effective source when $\Upsilon \rightarrow \ ^{c}\Upsilon $ with $\ ^{c}\Upsilon -4s^{2}\neq 0$ and the effective cosmological constant $\Lambda _{0}^{c}$ is chosen to satisfy the condition $\Lambda _{0}^{c}-4s^{2}\neq 0.$} We follow the same procedure as in sections \ref{sslccond} and \ref{ssecbcosm} and write in terms of
polarization functions,
\begin{eqnarray}
\ \ ^{c}\widehat{\mathbf{g}} &\mathbf{=}&\ ^{c}\eta _{1}\ ^{B}\mathring{g}%
_{1}dx^{1}{\otimes }dx^{1}+\ ^{c}\eta _{2}\ ^{B}\mathring{g}_{2}dx^{2}{%
\otimes }dx^{2}+(\ _{c}^{B}\omega )^{2}\left[ \ ^{c}\eta _{3}\ ^{B}\mathring{%
h}_{3}\ ^{c}\mathbf{e}^{3}{\otimes }\ ^{c}\mathbf{e}^{3}+\ ^{c}\eta _{4}\
^{B}\mathring{h}_{4}\ ^{c}\mathbf{e}^{4}{\otimes }\ ^{c}\mathbf{e}^{4}\right]
,  \notag \\
\ ^{c}\mathbf{e}^{3} &=&dy^{3}+(\ ^{B}\mathring{n}_{1}+\ ^{c}\eta
_{1}^{3})dx^{1}+(\ ^{B}\mathring{n}_{2}+\ ^{c}\eta _{2}^{3})dx^{2},
\label{bianchattr} \\
\ ^{c}\mathbf{e}^{4} &=&dt+(\ ^{B}\mathring{w}_{1}+\ ^{c}\eta
_{1}^{4})dx^{1}+(\ ^{B}\mathring{w}_{2}+\ ^{c}\eta _{2}^{4})dx^{2}.  \notag
\end{eqnarray}%
We use double left labeling with "B" and "c" in order to emphasize possible Bianchi anisotropic and attractor like behaviour of certain geometric/ physical objects. The off--diagonal deformations with $\partial _{t}\
^{c}h_{a}\neq 0,\partial _{t}\ ^{c}\varpi \neq 0$ are determined by
\begin{equation*}
\ _{c}^{B}g_{1}=\ ^{c}\eta _{1}\ ^{B}\mathring{g}_{1}=e^{\ ^{c}\psi
(x^{k})},\ \ _{c}^{B}g_{2}=\ ^{c}\eta _{2}\ ^{B}\mathring{g}_{2}=e^{\
^{c}\psi (x^{k})},
\end{equation*}%
for $\ ^{c}\psi (x^{k})$ being a solution of the Poisson equation $\partial
_{11}^{2}\ ^{c}\psi +\partial _{22}^{2}\ ^{c}\psi =2(\ ^{c}\Upsilon -4s^{2}), $ and

\begin{equation*}
\ _{c}^{B}h_{3}=\ ^{c}\eta _{3}\ ^{B}\mathring{h}_{3}=\frac{\ _{c}^{B}%
\widetilde{\Psi }^{2}}{4(\Lambda _{0}^{c}-4s^{2})}\mbox{ and
}\ _{c}^{B}h_{4}=\ ^{c}\eta _{4}\ ^{B}\mathring{h}_{4}=\frac{(\partial _{t}\
_{c}^{B}\widetilde{\Psi })^{2}}{\ _{c}^{B}\Xi }.
\end{equation*}%
The generating functions encode data on inhomogeneous locally anisotropic
interactions, attractor configurations and EYMH sourses,
\begin{equation*}
\ _{c}^{B}\Psi ^{2}=(\Lambda _{0}^{c}-4s^{2})^{-1}\int dt(\ ^{c}\Upsilon
-4s^{2})\partial _{t}(\ _{c}^{B}\widetilde{\Psi }^{2})\mbox{
or }\ _{c}^{B}\widetilde{\Psi }^{2}=(\Lambda _{0}^{c}-4s^{2})\int dt(\
^{c}\Upsilon -4s^{2})^{-1}\partial _{t}(\ _{c}^{B}\Psi ^{2})
\end{equation*}%
which results in a re--defined source , $\ ^{c}\Upsilon -4s^{2}\rightarrow \
^{B}\Xi ,$ with $\ ^{c}\Upsilon -4s^{2}=\partial _{t}(\ _{c}^{B}\Xi
)/\partial _{t}(\ _{c}^{B}\widetilde{\Psi }^{2}),$ when
\begin{equation*}
\ _{c}^{B}\Xi \lbrack \Upsilon ,\ _{c}^{B}\widetilde{\Psi }]=\int dt(\
^{c}\Upsilon -4s^{2})\partial _{t}(\ _{c}^{B}\widetilde{\Psi }^{2})
\end{equation*}
for a prescribed generating function $\ _{c}^{B}\widetilde{\Psi }.$ The N--connection coefficients in (\ref{bianchattr}) are computed
\begin{eqnarray*}
\ _{c}^{B}n_{k}(x^{k},t) &=&\ ^{B}\mathring{n}_{k}+\ ^{c}\eta _{k}^{3}=\
_{1}n_{k}(x^{i})+_{2}\widetilde{n}_{k}(x^{i})\int dt\frac{(\partial _{t}\
_{c}^{B}\widetilde{\Psi })^{2}}{(\ _{c}^{B}\widetilde{\Psi })^{3}\
_{c}^{B}\Xi }\mbox{
and } \\
\ _{c}^{B}w_{i}(x^{k},t) &=&\ ^{B}\mathring{w}_{i}+\ ^{c}\eta _{i}^{4}=\frac{%
\partial _{i}(\ _{c}^{B}\Xi )}{\partial _{t}(\ _{c}^{B}\Xi )}.
\end{eqnarray*}%
Following the procedure explained in section \ref{sslccond}, we  impose additional constraints and extract LC--configurations.

Dependency on all spacetime coordinates are modelled via a $v$--conformal factor $\ _{c}^{B}\omega (x^{k},y^{3},t)$ (in indirect form, it also contain attractor properties) as a solution of (\ref{confeq}) with
attractor coefficients stated above when
\begin{equation*}
\partial _{i}(\ _{c}^{B}\omega ) -\ _{c}^{B}n_{i}\ (\partial _{3}\
_{c}^{B}\omega) -\ _{c}^{B}w_{i}\ (\partial _{t}\ _{c}^{B}\omega) =0.
\end{equation*}%
Restricting the class of generating functions, we  extract homogeneous configurations but with anisotropies when parameterizations are of the type $\ _{c}^{B}\widetilde{\Psi }(t),\ ^{c}\Upsilon (t),\ _{c}^{B}h_{a}(t),$ $\
_{c}^{B}w_{i}(t)$ and constant values for $\ _{c}^{B}g_{k}$ and $\ _{c}^{B}n_{i}.$

Applying the AFDM, we  generate off--diagonal cosmological attractor
solutions of types 2 and 3  for the conventional and other families of inflation potentials,
for instance, when we use $\widetilde{q}(\frac{\phi /\sqrt{\ ^{c}\alpha }}{%
1+\phi /\sqrt{\ ^{c}\alpha }})$ instead of $q(\phi /\sqrt{\ ^{c}\alpha })$
 \cite{linde1}. We note that we have used a different
system of notations and  our approach is based on geometric methods
which allows us to construct exact solutions of modified gravitational and
matter field equations. For certain well defined conditions, we reproduce
the results and "diagonal" models studied in (supersymmetric) models with
dark matter and dark energy effects. Nevertheless, nonlinear parametric
systems of PDEs corresponding to effective EYMH interactions in (modified)
TMTs contain  solutions at a richer level  that were not analyzed and applied
to modern cosmology. Even though the off--diagonal effects at large observational
scales seem to be very small, the generic nonlinear character of
cosmological solutions depending on space like coordinates result in  new
nonlinear physics described by re--scaling via generating functions and
effective sources. Attractor type configurations offer alternative solutions  of crucial
importance for explaining the inflation scenarios in modern cosmology.
\vskip .3 in
\subsection{Cosmological implications of TMT nonholonomic attractor type
configurations}
\vskip .3 in
Here we concentrate on observational consequences of generic off--diagonal
solutions for  the effective EYMH systems with attractor properties in TMTs. We
have demonstrated  that Lagrangians of type $\ _{q}^{1}L$ (\ref{qth}) and $\ _{\chi
}^{1}L$ (\ref{eqth}) and their effective energy--momentum tensors are
naturally included as sources (\ref{effemt}) in action (\ref{acttmt}) with
two measures, which result in a nonholonomic modification of Einstein
gravity (\ref{effectconf}). Geometrically, we  reproduce such effects via
re--definition of generating functions (\ref{rescalling}) and fixing a
cut off constant $\ ^{c}\alpha $ for attractor configurations, when the
effective matter field interactions are modelled for a nonholonomic
off--diagonal vacuum configuration with small effective cosmological
constant and gravitational $\eta $--polarizations.

In general,  proposing and observing  physical realizations for
solutions with arbitrary $\eta $--deformations of well known prime
cosmological metrics (for instance, of FLRW, Bianchi or Kasner type ones) are difficult.
Nevertheless, we have elaborated upon the large distance inflationary scenarios if $%
\eta \approx 1+\varepsilon \widetilde{\eta }$ when $|\varepsilon \widetilde{%
\eta }|\ll 1.$ We note that such configurations encode nonlinear parametric
effects even when the off--diagonal and inhomogeneous terms are not taken into
consideration in order to explain certain observational data.
Using the results of analysis for $\ _{q}^{1}L$ (\ref{qth}) and
LC--configurations \cite{linde1}, we conclude and speculate on such
observational consequences:

\begin{enumerate}
\item TMTs and nonholonomic modifications of the EYMH theory contain inflationary model of the plateau--type and features of universal attractor property when $n_{s}=1-2/N$ and $r=12\ ^{c}\alpha /N^{2}.$

\item For $\ ^{c}\alpha =1$ such models are related to cosmological scenarios with the Starobinsky type model and Higgs inflation \cite{starob1,starob2,salop,bezr}; we obtain an asymptotic theory for  quadratic
inflation with $n_{s}=1-2/N,r=8/N,$ for large cutoff $\ ^{c}\alpha .$

\item Decreasing $\ ^{c}\alpha ,$ we get a universal attractor property both for diagonal and off--diagonal configurations; there are many models which have the same values $n_{s}$ and $r.$ This property is  preserved
for EYMH contributions, solitonic and/or gravitational waves for corresponding nonholonomic configurations.

\item In the limit of large $\ ^{c}\alpha ,$ we have  generated models of simplest chaotic inflation.  We have shown that effective nonlinear potentials with a second attractor are other viable possibilities.

\item For intermediate values of $\ ^{c}\alpha ,$ the predictions
interpolate between these two critical points, thus oscillating between  the sweet
spots of both Planck and BICEP2 \cite{planck1,planck2,bicep2}.
\end{enumerate}

With respect to the old and new cosmological problem, the issues 1-5 is analyzed in the context of TMTs when the constructions are naturally extended to include effective gauge field contributions which, in turn, modify nonlinearly the sources, effective cosmological constant and generating functions. Via conformal transforms, the attractor configurations are related to inhomogeneous and locally anisotropic solutions in modified gravity theories.  It is not surprising that the cosmological attractor configurations with TMT and nonhlonomic modifications of the EYMH theory are described in diagonal limits by the same parametric data as for the  holonomic attractor solutions \cite{linde1,linde2,linde3}.  We imposed such nonholonomic constraints and selected respective generating functions which reproduce this class of cosmological solutions. Nevertheless, the constants $\ ^c\alpha , n_s, r $ encode contributions from modified gravity theories and off--diagonal gravitational and matter field interactions and result in different observational consequences.

\section{Concluding Remarks}
\label{s6}
To mention a few, the most important physical solutions in modern gravity and cosmology theories  pertaining to black holes,  wormhole configurations, FLRW metrics, are  constructed for diagonal metrics transforming the
(modified) Einstein equations into certain nonlinear systems of second (or
higher) order ODEs. The solutions generally depend on integration constants. Such
constants are fixed following certain symmetry and other physical
assumptions in order to explain and describe the experimental and observational
data. There are also constructed more sophisticated classes of solutions,
for instance, with off--diagonal rotating metrics with Killing,  Lie type
symmetries and solitonic hierarchies which  provide important examples
of nonlinear models of gravitational and matter interactions.
Nevertheless, the bulk of such analytic and numerical methods of
constructing exact solutions are based on certain assumptions where the
corresponding nonlinear system of PDEs are transformed into ODEs. The
solutions are parameterized via integration parameters, symmetry and
physical constants. The main idea is to formulate an approach to simplify the equations and find solutions
depending, for instance, on a radial or a time like variable. The drawback of this approach is that a
number of nonlinear parametrical solutions are lost and thus unavailable  for
possible applications in cosmology and astrophysics.

The AFDM is presented as a geometric method for constructing general
classes of off--diagonal metrics, auxiliary connections and adapted frames
of reference when gravitational and matter field equations in various
modified/ generalized gravity theories, including general relativity,
are decoupled. This decoupling implies that the corresponding nonlinear
system of PDEs splits into certain subclasses of equations which contain
partial derivative depending only on one coordinate and relates only two unknown
variables and/or generating functions. As a result, we can integrate such
systems in very general off--diagonal forms when various classes of
solutions are determined not only by integration constants but also by
generating and integration functions, symmetry parameters and anholonomy
relations. The solutions depend, in general, on all spacetime
coordinates and can be with Killing or non--Killing symmetries, of different
smooth classes, with singularities and nontrivial topology. We can make, for
instance, certain approximations on the type of generating functions and
effective source at the end, after a general form of solution has been
constructed. This way we  generate new classes of cosmological metrics
which are  homogeneous or inhomogeneous, and in general, with local
anisotropies, which can not be found if one works from the very beginning with
simplified ansatz and higher symmetries. Furthermore, the possibility to re--define the
generating functions and sources via nonlinear frame transformations and
parametric deformations allows one to entertain  new classes of
solutions and study various nonlinear physical effects.

In this paper, we studied in explicit form certain classes of modified
gravity theories which can be modeled as TMTs with effective EYMH
interactions. Possible scalar fields and corresponding nonlinear interaction
potentials were chosen to select and reproduce attractor type solutions
with cut off constants which seem to have fundamental implication in
elaborating isotropic and anisotropic inflation scenarios in modern
cosmology. In general, one  can work with off--diagonal configurations and
consider diagonal limits for minimal and/or non--minimal coupling constants.
We proved that the decoupling property holds also in TMTs which results in
the possibility of constructing various classes of off--diagonal cosmological
solutions with small vacuum density. Such solutions describe spacetimes
with nonholonomically induced torsion. Nevertheless we formulated well--defined criteria
when additional nonholonomic constraints are introduced that allow to extract
LC--configurations. We studied nonholonomic deformations of FLRW, Bianchi
and Kasner type metrics encoding TMT effects and possible contributions of
effective EYMH interactions.

We have shown that attractor type cosmological solutions with cut--off parameters
can be derived by nonlinear re--definitions of
generating functions and effective sources in TMT if a corresponding type of
nonlinear scalar potential is chosen. In general, such attractor solutions
are model independent and  are constructed in explicit form to accommodate
effective EYMH interactions. In this way various large scale inflationary  models,
with anisotropic expansion and parametric nonlinear processes
can be realized.

For certain conditions, the gravitational and matter field equations of TMTs are expressed as effective Einstein equations with non-minimal coupling \cite{guend7}. In this presentation, we proved that in nonholonomic  N-adapted variables and for additional assumptions the constructions are  generalized in such form that two measure configurations  serve  to encode massive gravity effects and nonlinear parametric off-diagonal interactions (see formulas (\ref{effectconf})-(\ref {acttmt})). In general, such a theory also has 4 extra degrees of freedom with the Boulware-Deser (BD) ghosts. This problem can be circumvented if one imposes additional constraints. We imposed nonholonomic constraints for constructing cosmological attractor configurations.  This procedure constrains the extra dimension degrees of freedom and encodes the TMT and massive term contributions into certain subclasses of solutions for off-diagonal effective Einstein spaces (see similar constructions for ghost-free massive $f(R)$ theories in Refs.  \cite{veffmassiv1}).  We conclude that  in our models the BD ghosts are absent for such special classes of nonholonomic configurations if generic off-diagonal cosmological solutions are constructed for effective Einstein equations of type (\ref {eq1b})-(\ref {lccondb}).

There remain many open questions on how to provide viable explanations for
the recent observational data from Plank and BICEP. In this work, we
have shown that attractor configurations can be constructed in TMTs with
effective gravitational and matter field equations. Such solutions provide a new
background for investigating cosmological theories with anisotropies,
inhomogeneities, dark energy and dark matter physics.

\vskip5pt

\textbf{Acknowledgments:\ }    SV  reports certain research related to his former basic activity at UAIC; the Program IDEI, PN-II-ID-PCE-2011-3-0256;  a DAAD fellowship in 2015 and  support from Quantum Gravity Research, QGR-Topanga, California, USA.

\end{document}